\documentclass[preprint,aps,showpacs,prb,amsmath,amssymb,floatfix]{revtex4-1}         %preprint
\usepackage{longtable}
\usepackage{textcomp}
\usepackage{graphicx}% Include figure files
\usepackage{dcolumn}% Align table columns on decimal point
\usepackage{bm}% bold math
\usepackage{rotating}
\usepackage{color}
\usepackage{ulem}
\def\ket#1{|#1\rangle}                     % ket
\def\bra#1{\langle#1|}                     % bra
\def\skp#1#2{\langle#1|#2\rangle}          % scalar product
\usepackage{amsmath}

\begin{document}

%\preprint{Version 22 \date{\today}}
%

%
\title{Tuning the laser-induced ultrafast demagnetization of transition metals}

\author{W. T{\"o}ws}
\author{G. M. Pastor}
\affiliation{
Institut f{\"u}r Theoretische Physik,
Universit{\"a}t Kassel,
Heinrich-Plett-Stra{\ss}e 40, 34132 Kassel, Germany
}
\date{\today}

\begin{abstract}

The ultrafast demagnetization (UFD) dynamics of itinerant ferromagnets is theoretically investigated 
as a function of the characteristics of the initial laser excitation. A many-body $pd$-band 
Hamiltonian is considered which takes into 
account hybridizations, Coulomb interactions, spin-orbit interactions and the coupling to the laser field on the same 
electronic level. In this way, a fruitful connection is established between the non-adiabatic quantum
dynamics and the well-known equilibrium statistical mechanics of itinerant-electron ferromagnets.
The time evolution during and after the pulse absorption is determined exactly by performing 
numerical Lanczos propagations on a small cluster model with parameters appropriate for Ni. The most relevant 
laser parameters, namely, the fluence $F$, wave length $\lambda$, polarization $\hat \varepsilon$ 
and pulse duration $\tau_p$ are varied systematically. The results show how $F$, $\hat \varepsilon$ 
and $\tau_p$ allow one to control the 
total absorbed energy, the spectral distribution of the initial excitation, and the subsequent 
magnetization dynamics. The calculations show that reasonable changes in these parameters do not 
affect the UFD dynamics qualitatively and have only 
a minor influence 
on the time scale $\tau_\text{dm}$ which characterizes the initial demagnetization. In contrast, our 
model predicts that the degree of demagnetization 
$\Delta S_z / S_z^0$ achieved for $t \gtrsim \tau_\text{dm}$ correlates well with the 
average number of electrons excited by the laser or average number of absorbed photons 
$n_\text{ph}$, which can be tuned by varying the fluence, spectral distribution and 
polarization of the laser pulse.
The theoretical results are discussed by comparing them with available experiments.
From a fundamental perspective, the robustness of the ultrafast demagnetization effect is 
theoretically demonstrated, as a phenomenon reflecting the intrinsic dynamics of 
the metallic $3d$ valence electrons. A wide variety of well-focused possibilities of tailoring the 
efficacy of the ultrafast demagnetization process is thereby opened.

\end{abstract}
%

% Classification Scheme.
%\keywords{Suggested keywords}%Use show-keys class option if keyword
%                              %display desired
%
%\vspace{2mm}
%\pacs{%\\
%75.75.+a, %Magnetic properties of nanostructures\\
%36.40.Cg, %Electronic and magnetic properties of clusters\\ 
%75.50.Bb, %Fe and its alloys\\ 
%73.22.-f  %Electronic structure of nanoscale materials: 
%          %clusters, nanoparticles, nanotubes, and nanocrystal\\
%}
%
% 

\maketitle

\section{Introduction}
\label{sec:introduction}

Over the past twenty years, a wide range of different
time-resolved experiments have demonstrated that the 
excitation of 
magnetic transition metals (TMs) and rare earths with short laser pulses triggers an ultrafast demagnetization (UFD) of 
the material on a subpicosecond or picosecond 
time scale.\cite{Bea96,Hoh97,Koo00,Rhi03,Lis05,Cin06,Sta07,Car08,Mel08,Sta10,Koo10,Wie11,Car15,Eic17,
Ten18,Gor18} This remarkable effect offers new possibilities of ultrafast control and manipulation 
of the magnetization, which could find multiple applications in spin-electronic devices and storage 
media. Understanding the non-trivial quantum physics behind this phenomenon is obviously crucial for 
any knowledge-oriented material design. 
Therefore, several mechanisms explaining the UFD have been proposed in the 
literature.\cite{Car08,Koo10,Car15,Eic17,Kaz07,Atx07,Kra09,Sch10,Mue14,Lec17,Koo05,Koo05-PRL,Ste09,
Ste10,Fae11,Car11,Mue13,Bat10,Mel11,Bat12,Rud12,Esc13,Tur13,Sch13,Dew18,Zha00,Big09,Von12,Kri15,
Ell16,Kri17,Sho17,Toe15} On the one hand, one finds models in which the central role is played by 
the coupling between the narrow-band electrons responsible for magnetism and some distinct, a priori 
nonmagnetic degrees of freedom or field. In this context, let us mention the mechanisms based on 
electron-phonon spin-flip scattering,\cite{Koo10,Koo05,Koo05-PRL,Ste09,Ste10,Fae11,Car11,Mue13} on 
the transport of spin-polarized electrons,\cite{Bat10,Mel11,Bat12,Rud12,Esc13,Tur13,Sch13,Dew18} and 
on the coherent relativistic interaction between the photon field and the electronic 
spins.\cite{Zha00,Big09,Von12}
On the other hand, two purely electronic theories have also been 
proposed, in which the essential part of the demagnetization takes place within the electronic 
system, as a result of the coupling between the spin and translational degrees of freedom in the 
presence of the lattice potential.\cite{Kri15,Ell16,Kri17,Sho17,Toe15} To this category belong the 
time-dependent density-functional studies reported in Refs.~\onlinecite{Kri15,Ell16,Kri17,Sho17}. 
These explain the UFD as a throughout breakdown of the spin density and local 
magnetic moments in all unit cells, which involves spin-orbit driven spin flips and spin currents. 
An alternative approach, which is particularly relevant for the present paper, is the many-body 
electronic model Hamiltonian proposed in Ref.~\onlinecite{Toe15}. In this case the experimentally 
observed demagnetization is explained as the consequence of an ultrafast breakdown of the 
FM correlations between the local $3d$ magnetic moments which remain highly stable at all times.
From the latter investigations the following microscopic picture of the magnetization dynamics 
emerges:\cite{Toe15} (i)~At the start the laser excitation changes the occupation of the 
valence-electron 
states by inducing mainly $3d$-to-$4p$ electronic transitions, thus creating holes 
in the magnetically relevant $3d$ band. During this process the total 
magnetization of the sample remains essentially unchanged, since spin is conserved in optical 
transitions. (ii)~These 
changes in occupations trigger the dynamics by opening
so far Pauli-blocked new channels for spin-orbit coupling (SOC) induced
local angular-momentum transfers, dominantly from the atomic $d$-electron spins $\vec s_i$ to the 
$d$-orbital moments $\vec l_i$.
Taking into account that
the local spin moments are initially large and the orbital moments almost quenched, this process 
alone would tend to enhance $\langle {\vec l}_i \rangle$ at the expense of $\langle {\vec s}_i 
\rangle$,
since the total local angular momentum $\vec j_i = \vec l_i + \vec s_i$ is 
conserved by the SOC.
(iii)~However, the perpetual motion of electrons in the lattice due to 
interatomic hopping quenches
most efficiently any
incipient increase of the average orbital angular momentum $\langle \vec L \rangle = \sum_i \langle 
\vec l_i \rangle$ on a time scale of the order of one femtosecond. The result of these three simple 
fundamental processes is the rapid decrease of the average
electronic angular momentum $\langle \vec J \rangle = \sum_i \langle \vec j_i \rangle$ 
and magnetization of the sample as a function of time. The demagnetization occurs essentially at the 
same rate as the 
spin-to-orbital angular momentum transfer, which is governed by the SOC and thus corresponds to
a characteristic demagnetization time $\tau_\text{dm}$ of the order of $100$~fs.
Notice, moreover, that the sum of the angular momenta associated to the electronic 
and ionic degrees of freedom is strictly preserved by the electron-lattice interactions. Therefore, 
the decrease of $\langle \vec J \rangle$ is exactly compensated by an increase of the lattice 
angular momentum $\vec L_\text{lat}$, occurring at the same rate. The fact that high local-moment 
stability, electron delocalization, and spin-orbit interactions 
are all inherent features of itinerant-electron magnetism explains the experimentally observed 
universality of 
the UFD effect. Further details on the electronic mechanism of UFD are discussed in 
Ref.~\onlinecite{Toe15}.

In past years a considerable research activity has been focused on the role of the initial 
excitation in the UFD process, and on the possibilities of controlling the spin dynamics by tuning 
the laser-pulse characteristics.\cite{Kho12,Ost12,Ell16,Bie17} For example, it has been recently 
demonstrated that the degree of 
demagnetization can in principle be controlled 
by changing the shape and spectral distribution of the pump pulse.\cite{Ell16}
It is therefore 
most interesting to correlate the degree of demagnetization with the material parameters and 
electronic structure. Furthermore, one would like to understand how the efficiency of the 
demagnetization process depends on the degree of excitation of the ferromagnet. Varying
the intensity of the pumping pulse at a given frequency allows us to adjust the number of absorbed 
photons, excited electrons and absorbed energy. Changing the laser frequency for a given absorbed 
energy one should be able to discern the role of the number of excitations, and 
thus gain further insight into thermalization effects. In addition, one may also 
consider different circularly and linearly polarized light, in order to explore how an initial 
transfer of angular momentum upon laser absorption may affect the subsequent dynamics. Finally, 
adjusting the laser-pulse duration $\tau_p$, from 
ultrashort highly-intense excitations to values of $\tau_p$ comparable with SOC relaxation time, 
should help us to reveal any specific spin dynamics taking place while the 
laser field is active, and which may result from SOC-laser interference 
effects.\cite{Big09,Von12,Zha00}
It is the purpose of this paper to investigate the role of the initial laser excitation 
on the magnetization dynamics of ferromagnetic TMs and to quantify the possibilities of tuning the 
ultrafast demagnetization by its means. To this aim we consider a many-body electronic theory in 
which the dynamics of the electronic translational, orbital 
and spin degrees of freedom, as well as their coupling to the external electric field, are described 
quantum mechanically and on the same footing.\cite{Toe15}

The remainder of the paper is organized as 
follows. The theoretical background, including a derivation of the model Hamiltonian, the 
involved approximations, and the parameters used for the calculations, is presented 
in Sec.~\ref{sec:model}. Exact numerical results for the magnetization dynamics as a function of the 
fluence, wave length, polarization and duration of the laser pulse are presented and discussed in 
Sec.~\ref{sec:pulse-parameters}. Finally, Sec.~\ref{sec:conclusion} summarizes the main conclusions 
and perspectives.

\section{Theoretical background}
\label{sec:model}

In the following we first derive the electronic model\cite{Toe15} used in the present 
investigations of the laser-induced magnetization dynamics by explicitly pointing out all the 
involved approximations. The complete many-body problem, which includes both electronic and ionic 
degrees of freedom, is simplified by using the \textit{Born Oppenheimer} approximation, which 
decouples the electronic and ionic dynamics.\cite{Bor27} This is justified, as usual, by the large 
ion-to-electron mass ratio, and the resulting differences in the corresponding time scales. Since we 
are interested in the dynamics of the magnetization, which is given by the spin and orbital 
electronic 
contributions, we focus on the electronic degrees of freedom so that the ion coordinates 
appear only as parameters of the quantum many-electron problem. Although the lattice dynamics is 
ignored in all the calculations reported in Sec.~\ref{sec:pulse-parameters}, we shall 
return to it at the end of this section and in Sec.~\ref{sec:conclusion}, when discussing the 
conservation of total (lattice plus electron) angular momentum and the possible role of the coupling 
to phonons.

The spin and orbital magnetic moments of transition metals are known to be dominated by the 
$3d$-electron contributions. Moreover, the prime optical excitations of the $3d$ states, which 
result from the pumping laser, involve transitions to the nearby $4p$ orbitals. Therefore, in order 
to capture the main physics of laser-excited $3d$ electrons in ferromagnetic TMs, it is reasonable 
to concentrate on the correlated-electron dynamics within these two bands. The corresponding 
many-body $pd$ Hamiltonian is given by
\begin{equation}\label{eq:modelham}
  \hat H = \hat H_0 + \hat H_{C} + \hat H_\text{SO} + \hat H_{E}(t) ~,
\end{equation}
where
\begin{equation}\label{eq:tightbinding}
  \hat H_0 = \sum_{i \alpha \sigma} \varepsilon_\alpha \, \hat n_{i \alpha \sigma} + \sum_{ij} 
\sum_{\alpha \beta 
\sigma} 
\, t_{ij}^{\alpha \beta} \, \hat c_{i\alpha\sigma}^\dagger \hat c_{j\beta\sigma}
\end{equation}
describes the single-particle electronic structure of the $3d$ and $4p$ bands. In the usual 
notation, $\hat c_{i\alpha\sigma}^\dagger$ ($\hat c_{i\alpha\sigma}$) creates (annihilates) 
an electron at atom $i$ with radial and orbital quantum numbers $\alpha = nlm$ and spin $\sigma$ 
($nl$ refers to $3d$ and $4p$). The corresponding electron number 
operator is $\hat n_{i \alpha \sigma}$. For simplicity, the energy levels $\varepsilon_\alpha$ of 
the atomic-like $3d$ and $4p$ orbitals $\ket{\varphi_{i\alpha}}$ are assumed to be 
independent of $m$.
The interatomic hopping integrals $t_{ij}^{\alpha\beta}$ describe the delocalization of the 
electrons throughout the lattice. Formally, they are given by $t_{ij}^{\alpha\beta} = 
\bra{\varphi_{i\alpha}} ( -{\hbar^2 \nabla^2}/{2\mu} + \phi^\text{lat} ) 
\ket{\varphi_{j\beta}}$, where $\mu$ stands for the electron mass and $\phi^\text{lat}$ for the 
effective lattice potential, which depends on all atomic positions $\vec R_i$. 
Notice that the hoppings $t_{ij}^{\alpha\beta}$, but also the energy levels $\varepsilon_\alpha = 
t_{ii}^{\alpha\alpha}$, incorporate the leading contribution to the electron dynamics resulting 
from the electron-lattice interaction as given by $\phi^\text{lat}$. In the following, the 
hopping integrals $t_{ij}^{\alpha\beta}$ are determined by using the two-center approximation, which 
takes into account the most important terms in $\phi^\text{lat}$ due to the ions $i$ and 
$j$.\cite{Sla54} In this case $t_{ij}^{\alpha\beta}$ depends only on the 
relative vector $\vec R_{ij} = \vec R_i - \vec R_j$, as well as on the radial and orbital quantum 
numbers $nlm$ of the orbitals $\alpha$ and $\beta$. Further details on the calculation of 
$t_{ij}^{\alpha\beta}$ may be found in Appendix~\ref{app:hopping}.

The second term, $\hat H_C$ in Eq.~\eqref{eq:modelham}, refers to the electron-electron interaction. 
For simplicity, we approximate it by taking into account only the dominant intra-atomic terms among 
the $3d$ electrons, which are known to be responsible for the magnetic behavior of 
TMs. Starting from the general intra-atomic expression
\begin{equation}
  \hat H_C = \dfrac{1}{2} \sum_{i} \sum_{\alpha \beta \gamma \delta \in 3d} 
\sum_{\sigma \sigma'} \, V_{\alpha \beta \gamma \delta} \, \hat c_{i\alpha\sigma}^\dagger \,
\hat c_{i\beta\sigma'}^\dagger \, \hat c_{i\delta\sigma'} \, \hat c_{i\gamma\sigma} ~,
\end{equation}
we consider only the largest two-orbital integrals, namely, the direct terms $U_{\alpha\beta} = 
V_{\alpha \beta \alpha \beta}$ and the exchange terms $J_{\alpha\beta} = V_{\alpha \beta \beta 
\alpha}$ ($\alpha \neq \beta$), which are the most important for the magnetic behavior. Moreover, 
the 
orbital dependences of $U_{\alpha \beta}$ and $J_{\alpha \beta}$ are neglected by setting them equal 
to their average values $U_{\alpha\beta} = U$ and $J_{\alpha\beta} = J$. While the orbital 
dependences of the intra-atomic $d$-electron repulsions are known to be important for a 
quantitative description of orbital magnetism,\cite{Nic06} they are not essential for describing 
the 
total spin 
polarization within the $3d$ band, even as a function of temperature.\cite{Gar15} Taking into 
account these 
simplifications one obtains the particularly transparent form\cite{Gar15,Uch01,Kak08,Kak11}
\begin{equation}
  \hat H_C \, = \, \dfrac{1}{2} \left(U - \dfrac{J}{2}\right) \sum_i \hat n_i^d \, \left( \hat n_i^d 
- 1 \right) \,-\, J \sum_i \hat{\vec s}_{i}^{\,d} \cdot \hat{\vec s}_{i}^{\,d} \,+\, \dfrac{J}{2} 
\sum_{i\alpha \in 3d} \hat n_{i\alpha} \left( 2 - \hat n_{i\alpha} \right) \, + \, \dfrac{J}{4} 
\sum_i \hat n_i^d ~.
\end{equation}
Here, $\hat n_i^d = \sum_{\alpha \in 3d, \sigma} \hat n_{i\alpha \sigma}$ denotes the operator for 
the total number of $3d$ electrons at atom $i$, $\hat{\vec s}_{i}^{\,\alpha}$ ($\hat n_{i\alpha}$) 
is the spin (number) operator for the orbital $\alpha$ at atom $i$, and 
$\hat{\vec s}_{i}^{\,d} = \sum_{\alpha \in 3d} \hat{\vec s}_{i}^{\,\alpha}$ the total $3d$-electron 
spin operator at atom $i$. The first terms, proportional to the number of pairs of $d$ electrons, 
take into account the changes in the Coulomb energy resulting from  charge fluctuations. Although 
important for correlations, they have a visibly non-magnetic character. The second terms, 
proportional to $(\hat{\vec s}_{i}^{\,d})^2$, favor a parallel alignment of all the $3d$ spins at 
each atom (Hund's first rule). They are responsible for the formation and strong stability of the 
local spin moments ($J \sim 1$~eV). Part of the energy gain upon local moment formation 
($33$--$50$\% depending on the number of unpaired electrons) is compensated by the third terms, 
which 
are proportional to $\hat n_{i\alpha} ( 2 - \hat n_{i\alpha})$. These terms are actually ignored in 
the subsequent dynamics, since their contribution results in an effective reduction of the 
exchange integral $J$, and since they are unaffected by the relative orientation of the 
unpaired spins. Finally, the last terms amount to an unimportant constant energy shift which can 
be 
incorporated in the definition of the bare levels $\varepsilon_{3d}$ [see 
Eq.~\eqref{eq:tightbinding}].

For the sake of compactness it is useful to define a new direct Coulomb repulsion parameter $U$ as 
the average repulsion $U - J/2$ between $d$ electrons having parallel and antiparallel spins. In 
this way one obtains the model interaction in its final form\cite{Toe15}
\begin{equation}\label{eq:interaction}
  \hat H_{C} = \dfrac{U}{2} \sum_i \hat n_i^d ( \hat n_i^d - 1) 
  - J \sum_i \hat{\vec s}_i^{\,\, d} \cdot \hat{\vec s}_i^{\,\, d} \, .
\end{equation}
Notice that $\hat H_C$, as the full Coulomb interaction, conserves both the spin $\vec s_i^{\,\,d}$ 
and orbital $\vec l_i$ angular momenta of the atoms, since the rotational invariance of the 
\textit{first-principles} interaction is not altered by the local approximations. In this context it 
is useful to recall that this model has been successfully applied in numerous previous studies of 
the equilibrium ground-state and finite-temperature properties of transition-metal 
magnetism.\cite{footnote1}

The third term in Eq.~\eqref{eq:modelham} is the spin-orbit coupling operator
\begin{equation}\label{eq:soc}
  \hat H_\text{SO} = {\xi} \, \sum_{i} \sum_{\alpha \alpha' \in 3d} \sum_{\sigma \sigma'} \, 
( \vec l 
\cdot \vec s )_{\alpha \sigma , \alpha' \sigma'} \, \hat c_{i \alpha \sigma}^\dagger \hat c_{i 
\alpha' \sigma'}
\end{equation}
in an intra-atomic approximation within the $3d$ band, where the parameter $\xi$ denotes the SOC 
strength. The matrix elements $( \vec l \cdot \vec s)_{\alpha \sigma , \alpha' \sigma'}$ of 
$\hat{\vec l}_i \cdot \hat{\vec s}_i$ at atom $i$ couple 
the spin and orbital degrees of freedom, thereby conserving the total local angular momentum $\vec 
j_i = \vec l_i + \vec s_i$.

The last term $\hat H_E$ in Eq.~\eqref{eq:modelham} introduces the interaction with the external 
laser field, which
is treated in the intra-atomic dipole approximation. 
For linearly polarized light we have
\begin{equation}\label{eq:dipole-approx}
  \hat H_E(t) = e \hat{\vec r} \cdot \vec E(t) = e |\vec E(t)| \sum_{i\alpha\beta\sigma} \, 
\bra{\alpha} 
\hat\varepsilon \cdot \hat {\vec r} \ket{\beta} \, \hat c_{i\alpha\sigma}^\dagger \hat 
c_{i\beta\sigma} ~,
\end{equation}
where $\vec E(t)$ refers to the uniform classical electric field,
$\hat \varepsilon$ denotes a dimensionless normalized polarization vector, and $e > 0$ is the 
electron charge. In the 
case of circularly polarized laser pulses $\hat H_E$ is replaced
by the operator $\hat H_E^\sigma$, which describes 
an electric field with helicity $\sigma = \pm 1$ carrying an angular momentum $\sigma \hbar$ along 
the $z$ axis. 
This is given by
\begin{equation}\label{eq:el-field-ham-circular}
  \hat H_E^{\pm} (t) ~=~ e | \vec E (t) | \, \hat P_p \, ( \hat \varepsilon_{\pm} \cdot \hat{\vec r} 
) \, \hat P_d \, + 
\, h.c. ~,
\end{equation}
where $\hat P_d$ ($\hat P_p$) denotes the projection operator onto the $3d$ ($4p$) orbitals and 
$\hat \varepsilon_{\pm} 
= ( \hat e_x \, \pm \, i \hat e_y ) / \sqrt{2}$ is the complex polarization vector. As usual, $\hat 
e_x$ and $\hat 
e_y$ stand for the unit vectors along the $x$ and $y$ axis. Since the dipole matrix elements 
$\bra{\alpha} \hat{\vec r} 
\ket{\beta}$ satisfy the atomic selection rule $\bra{nlm} \hat{\vec r} \ket{n'l'm'} = 0$ unless $l - 
l' = \pm 1$, the optical excitation involves only $3d$-$4p$ transitions. A more detailed account 
of the dipole 
matrix elements is 
given in Appendix~\ref{app:dipole}. The operator $\hat H_E^{+}$ can be interpreted as follows. The 
first term in Eq.~\eqref{eq:el-field-ham-circular}
describes the absorption of a photon which transfers an angular momentum $+ 
\hbar$ to a $3d$ electron making a transition to a $4p$ orbital ($m \to m+1$). Hermiticity, as 
ensured by the second 
term in Eq.~\eqref{eq:el-field-ham-circular}, implies the emission of a photon with angular momentum 
$\hbar$ in the 
reverse electronic transition from a $4p$ to a $3d$ orbital ($m \to m-1$). Analogously, the operator 
$\hat 
H_E^{-}$ with the 
opposite helicity $\sigma = -1$ describes the absorption (emission) of an angular momentum $-\hbar$ 
in the optical 
transitions from $3d$ to $4p$ ($4p$ to $3d$) orbitals.

Before closing the discussion of the model, it is worth recalling that the 
field-free Hamiltonian $\hat H = \hat H_0 + \hat H_C + \hat H_\text{SO}$ represents a 
purely electronic model, which describes the dynamics of electrons within the lattice potential 
$\phi^\text{lat}$ generated by the ions at given fixed positions $\vec R_j$. 
Since $\phi^\text{lat}$ is obviously not isotropic, the electronic angular momentum $\vec L + \vec 
S$ is not conserved, where $\vec L = \sum_i \vec l_i$ ($\vec S = \sum_i \vec s_i$) stands for the 
total electronic orbital (spin) angular momentum. However, the combined system of electrons and ions 
represents a closed and therefore rotationally invariant system. 
Consequently, it is clear that the total angular momentum of electrons and ions $\vec J = \vec L + 
\vec S + \vec L^\text{lat}$ remains a formally rigorous constant of motion, where $\vec 
L^\text{lat}$ denotes the angular momentum of the lattice. An explicit account of the time 
dependence of the lattice angular momentum would require to consider the dynamics of the ionic 
degrees of freedom, which is beyond the scope of the present work.

\subsection{Model simplifications and parameters}
\label{sec:application}

In order to achieve an exact numerical solution of 
the many-body dynamics without involving often hardly 
controllable and symmetry breaking mean-field approximations, we keep the $pd$-band model 
as transparent as possible by introducing two simplifications. First, we 
reduce the orbital 
degeneracy by considering only the $3d$ orbitals having $|m| \leq 1$ and the $4p$ orbital having $m = 
0$. This 
approximation reduces the numerical effort involved in the exact numerical propagation without 
affecting significantly 
the $3d$-$4p$ optical absorption process, the electronic delocalization and exchange interactions 
responsible for 
magnetism, or the angular-momentum transfer between spin and orbital degrees of freedom induced by 
the SOC. Similar 
reductions of the local orbital degeneracy have often been used in the context of electron 
correlations and itinerant 
magnetism, in particular, in connection with the Hubbard model.\cite{Hub63,Kan63,Gut63} 

The second approximation consists in performing the numerical propagations on a small cluster model. 
In this work, we 
consider equilateral triangles ($N_a = 3$ atoms) having $N_e = 4$, $5$ and $7$ electrons, and an 
equilateral $N_a = 4$ 
rhombus with $N_e = 5$, where the length of the short diagonal equals the side length. This allows us 
to explore 
various geometries and band fillings having different 
electronic structures. As we shall see in Sec.~\ref{sec:pulse-parameters}, the validity of this 
approximation 
can be justified \textit{a posteriori} by the local character of the mechanism responsible for 
angular 
momentum transfer and ultrafast demagnetization.

The model parameters are specified as follows. The hopping integrals $t_{jk}^{\alpha \beta}$ are 
determined by 
considering nearest-neighbor (NN) Slater-Koster integrals $(dd\sigma) = 0.6$~eV, 
$(dd\pi) = -0.3$~eV, $(pp\sigma) = 1.5$~eV and $(pd\sigma) = -0.4$~eV. These values are similar to 
those obtained in Ref.~\onlinecite{Papaconstantopoulos} by 
fitting the 
experimental Ni band structure. Notice that the largest $pp$-integral is roughly three times larger 
than the 
$dd$-integrals. This corresponds to a rather broad $sp$-band and a narrow $3d$-band, as found in $3d$ 
TMs such as Ni. 
Moreover, the NN $pd$-hopping $(pd\sigma) = -0.4$~eV is also considerably large, almost of the same 
order of magnitude 
as the difference between the $4p$ and the $3d$ energy levels
$\varepsilon_p - \varepsilon_d = 1.4$~eV.
This leads to a 
significant $pd$-hybridization and a small but not negligible $p$-level ground-state occupation, 
which is consistent with the $spd$-hybridization found in $3d$ TMs.

The largest energy scale is given by the direct Coulomb 
integral $U = 8.0$~eV, a value taken from experimental photoemission spectra and theoretical 
calculations 
of the Ni density of electron states.\cite{Hue75,Ebe80,Fel80,Vic85} The intra-atomic exchange 
integral $J$ yields 
stable FM ground states whose easy magnetization direction defines the $z$ axis. For the 
rhombus having $N_a = 4$ atoms and $N_e = 5$ electrons, we use $J = 1.5$~eV and obtain a ground-state 
off-plane 
spin polarization $S_z^0 = 2.15 \, \hbar$. For the 
triangular clusters having $N_e = 4$, $5$ and $7$ we use $J = 0.8$--$1.0$~eV. The ground states of 
the 
$N_e = 4$ and $5$ triangles exhibit off-plane easy magnetization 
axes and spin polarizations of $S_z^0 = 1.96 \, \hbar$ and $S_z^0 = 1.34 \, \hbar$, while 
the ground state of the $N_e = 7$ triangle exhibits an in-plane easy magnetization plane and
$S_z^0 = 1.52 \, \hbar$.

The smallest energy scale in the model is the spin-orbit 
coupling strength $\xi = -80$~meV. Typical values for $3d$ TMs are in the range $|\xi| = 
50$--$100$~meV.\cite{Bru93} 
Notice that the sign of $\xi$ has been changed for systems having a less than half-filled $d$ band, 
in order to reproduce the parallel alignment between $\vec L$ and $\vec S$ found in Ni, Co and 
Fe.\cite{Landau} This corresponds to performing the electron-hole transformation $\hat 
h_{i\alpha\sigma} = \hat c^\dagger_{i\alpha\sigma}$, which does not affect the Coulomb interaction 
and only changes the sign of the hopping integrals. Explicit calculations show that
changing the sign of $\xi$ does not affect the time dependence of the discussed
observables in any significant way.

The spin dynamics is triggered by an optical pump pulse having a Gaussian form
\begin{equation}\label{eq:el-field}
  \vec E (t) = \hat \varepsilon \cdot E_0 \cos (\omega t) \exp(-{t^2}/{\tau_p^2}) ~,
\end{equation}
where $\omega = 2 \pi c / \lambda$ is the laser frequency. The pulse, centered at $t = 0$, has 
a duration characterized by the pulse width $\tau_p$. The intensity of the electric field can be 
measured by the 
maximal amplitude $E_0$ of $\vec E(t)$, which is related to the energy flow per unit area or fluence 
$F$: $E_0 = ( {2}/{\pi} )^{1/4} \sqrt{{2F}/(c \varepsilon_0 \tau_p)}$, where $\varepsilon_0$ is the 
vacuum permittivity. In 
order to investigate the role of the pump-pulse parameters in the laser-induced magnetization 
dynamics, we vary 
$F$, $\hat \varepsilon$ and $\tau_p$ systematically. In this way we quantify the dependence of the 
spin 
relaxation on the initial laser excitation. In cases where $F$, $\tau_p$ or $\hat \varepsilon$ are 
not 
explicitly mentioned, we use $F = 40 ~\text{mJ/cm}^2$, $\tau_p = 5$~fs and a linear in-plane 
polarization $\hat \varepsilon$ along one NN bond in the triangle or along the long diagonal in the 
rhombus.
These reference values 
correspond approximately to the typical numbers of absorbed photons per atom and pulse durations 
considered in 
experiments.\cite{Rhi03,Tur13}
Finally, the strength of the coupling between the electronic degrees of freedom and 
$\vec E$ is characterized by the 
reduced matrix element $\bra{3d} |\hat T^{(1)}|\ket{4p}$ [see Eqs.~\eqref{eq:dipole_simplified} 
and~\eqref{eq:sigma1_circular} in Appendix~\ref{app:dipole}]. For the calculations throughout this 
work we set 
$\bra{3d}|\hat T^{(1)}|\ket{4p} = 0.5$\AA{}, which corresponds to the typical extension of $3d$ and 
$4s$ orbitals in 
$3d$ TMs. The precise value of $\bra{3d} |\hat T^{(1)}|\ket{4p}$ is not important for our 
conclusions.

In the following section we investigate the consequences of the laser excitation on the FM ground 
state 
$\ket{\Psi_0}$ by 
propagating $\ket{\Psi (t)}$ numerically under the action of the time-dependent electric 
field. The time evolution is calculated by using the short-time iterative Lanczos propagation 
method.\cite{Tan07} Once 
the many-body wave function $\ket{\Psi(t)}$ is obtained we compute the expectation values $O(t) = 
\bra{\Psi(t)} 
\hat O \ket{\Psi(t)}$ of the observables $\hat O$ of physical interest, for example, the total spin 
magnetization 
$\hat S_z$, the local spin and orbital moments $\hat{\vec s}_i$ and $\hat{\vec l}_i$, and the 
spin-correlation 
functions $\hat{\vec s}_i \cdot \hat{\vec s}_j$.

\section{Results and discussion}
\label{sec:pulse-parameters}

Before solving and analyzing the dynamics it is important to keep in mind that the hybridizations 
due to the electron-lattice interaction, the Coulomb interactions 
and the laser-absorption processes, which are described by $\hat H_0$, $\hat H_C$ and $\hat H_E$, 
all conserve the total spin ${\vec S} = \sum_i {\vec s}_i$, i.e., $[\hat H_0 , \hat{\vec{S}}] = 
[\hat H_C , \hat{\vec{S}}] = [\hat H_E 
,\hat{\vec{S}}] = 0$. The spin-rotational invariance is broken only by the SOC since $[\hat H_\text{SO} 
,\hat{\vec{S}}] \neq 0$. However, the SOC operator $\hat H_\text{SO}$ commutes with the sum $\hat{\vec l}_i + \hat{\vec 
s}_i$ of the local orbital and spin angular momenta at each TM atom. Therefore, any spin-flip 
process induced by SOC 
necessarily involves a local angular momentum transfer between $\vec s_i$ and $\vec l_i$, in which 
the sum $\vec l_i + 
\vec s_i$ is conserved. This local intra-atomic symmetry notwithstanding, neither the local orbital moment $\vec l_i$ 
nor the total orbital angular momentum $\vec L = \sum_i \vec l_i$ are conserved 
throughout the dynamics, since the lattice potential is not rotationally invariant (i.e., $[ \hat H_0 , 
\hat{\vec{l}}_i] \neq 0$ and $[ \hat H_0 , \hat{\vec{L}}] \neq 0$). This can be traced back to the fact that
the interatomic hoppings $t_{ij}^{\alpha \beta}$ connect orbitals with different 
azimuthal quantum numbers $m$ at different atoms. The previous fundamental symmetry considerations 
are essential for understanding the ultrafast magnetization 
dynamics from a microscopic quantum perspective. The use of time-dependent mean-field approximations to the dynamics 
seems very questionable in this context, because they 
artificially break the spin-rotational invariance with respect to $\hat H_0$, $\hat H_C$ and $\hat H_E$. In contrast, 
exact time propagations ---although limited in their application to small finite systems--- have the clear 
advantage of complying with all 
fundamental conservation laws. They should therefore allow us to derive rigorous conclusions.\cite{Toe15}

The purpose of this Section is to investigate the 
dynamics of ferromagnetic TMs as a function of the laser fluence $F$, photon 
energy $\hbar \omega$, electric-field polarization $\hat \varepsilon$ and pulse duration $\tau_p$, 
in order to 
quantify to what extent these experimentally tunable parameters can be used to taylor the 
magnetization dynamics. Results for different model systems and band fillings 
are contrasted. The correlations between degree of initial electronic excitation, absorbed energy, 
demagnetization time and degree of demagnetization are analyzed. General trends are inferred.

\subsection{Laser fluence}
\label{subsec:fluence}

\begin{figure}[t]
 \centering
 \includegraphics[width=7.6cm,keepaspectratio=true]{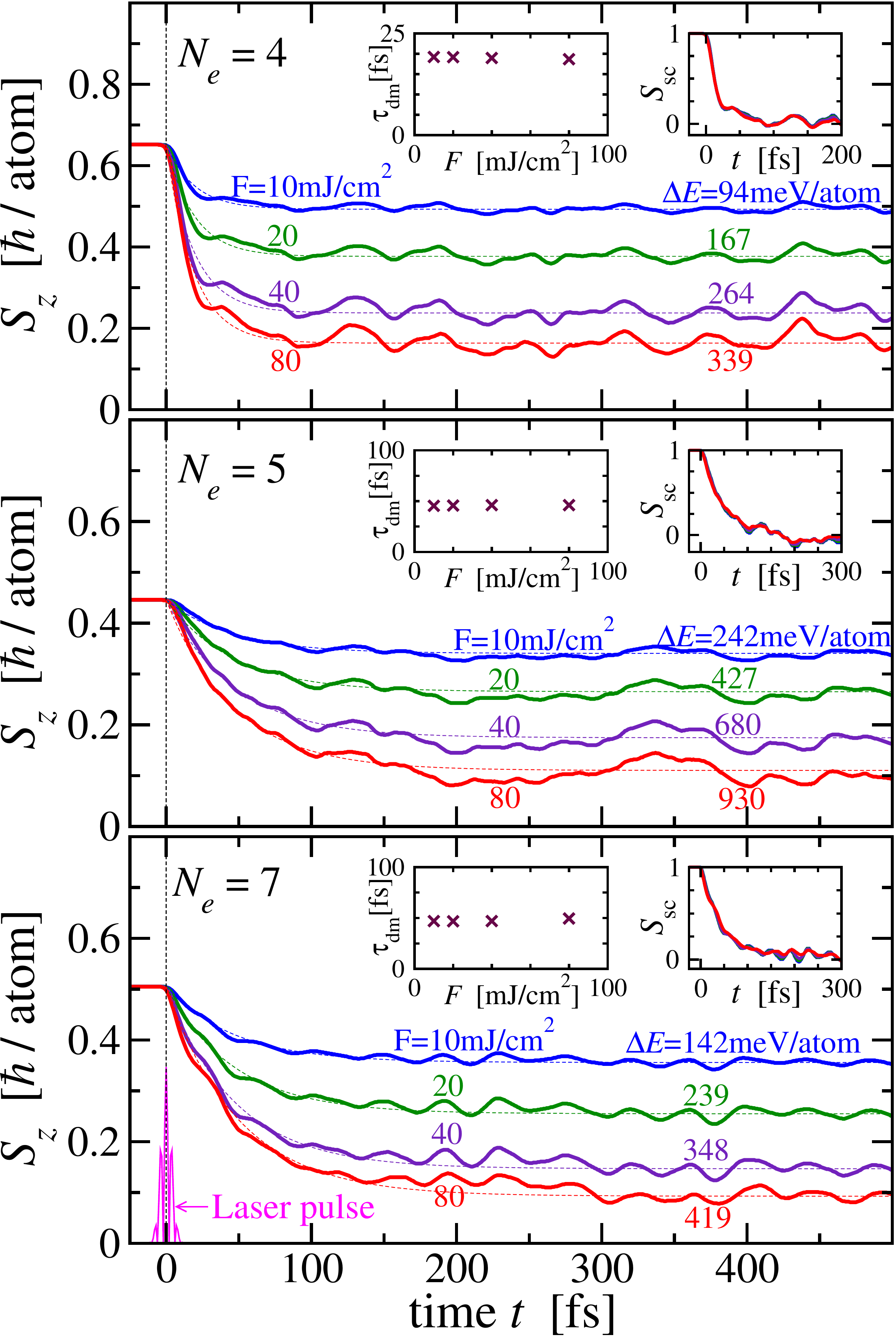} %width=8.663cm
 % figure1.eps: 0x0 pixel, 300dpi, 0.00x0.00 cm, bb=(atend)
 \caption{\small Time dependence of the spin magnetization in an equilateral triangle with $N_e = 4$, $5$ and $7$ 
electrons, after excitation using a linearly polarized $5$~fs laser pulse with wave length $\lambda 
= 1051$~nm, $556$~nm 
and $849$~nm, respectively. The considered laser fluences $F$ are indicated together with the corresponding absorbed 
energies per atom $\Delta E$. The left insets show the demagnetization times $\tau_\text{dm}$ as a function of $F$, as 
obtained from exponential fits to $S_z(t)$ given by the dashed curves in the main panels. The right 
insets show, for all considered values of $F$, the nearly identical time dependences of the 
corresponding scaled spin magnetization $S_\text{sc}(t) = 
[S_z(t) - S_z^\infty]/\Delta 
S_z$. In the bottom panel the amplitude $E(t)$ of the triggering electric field is illustrated.}
 \label{fig:figure1}
\end{figure}

\begin{figure}[t]
 \centering
 \includegraphics[width=7.6cm,keepaspectratio=true]{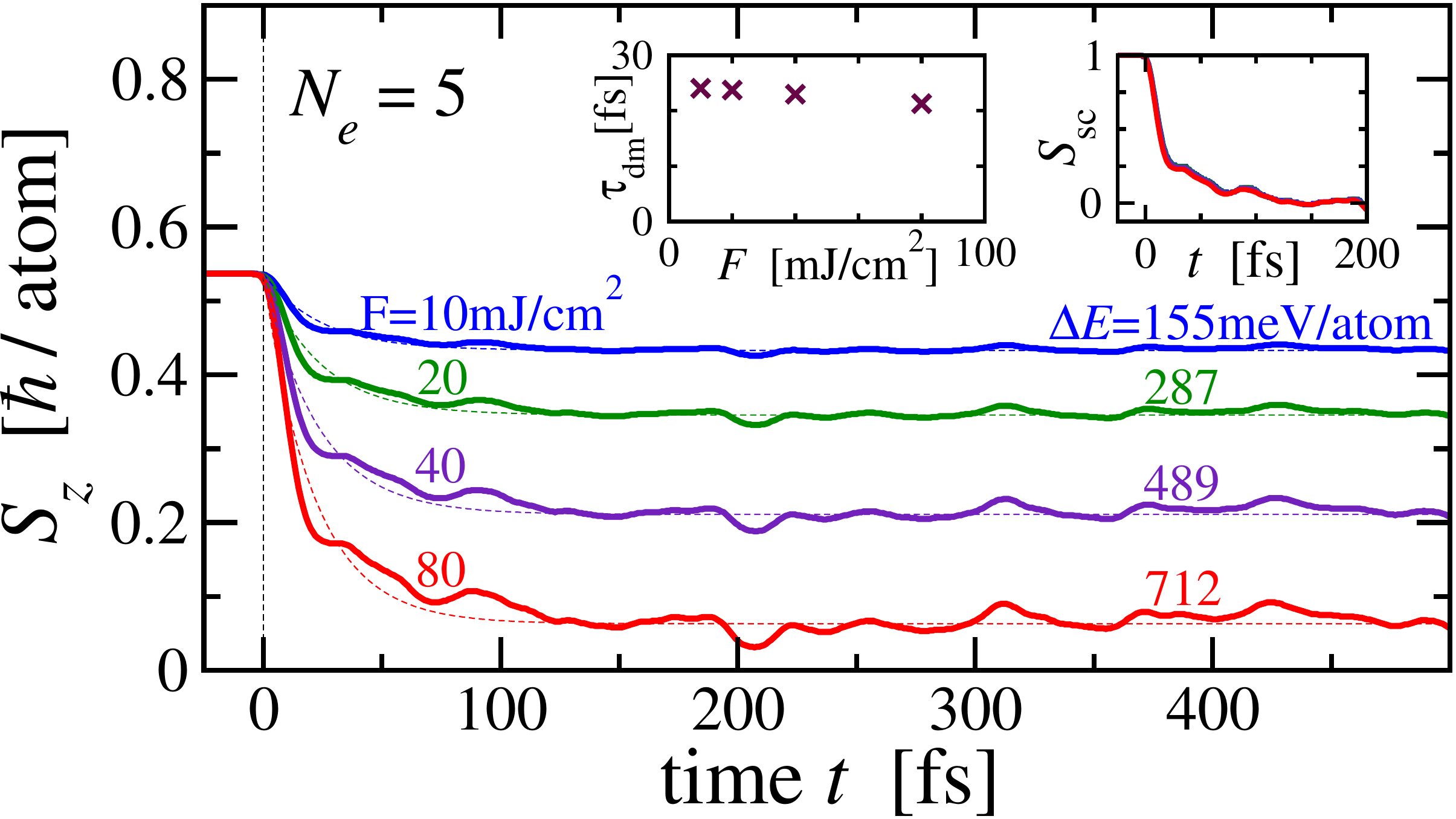} %width=8.663cm
 % figure1.eps: 0x0 pixel, 300dpi, 0.00x0.00 cm, bb=(atend)
 \caption{\small Time dependence of the spin magnetization in a rhombus having $N_e = 5$ 
valence electrons. The excitation at $t=0$ corresponds to a 
linearly polarized $5$~fs laser pulse with wave length $\lambda = 385$~nm. See the caption of 
Fig.~\ref{fig:figure1}.}
 \label{fig:figure2}
\end{figure}

The laser fluence $F$ is naturally expected to play an important role in the subsequent 
spin relaxation since it 
controls the level of electronic excitation. In order to quantify its effect we have determined the magnetization 
dynamics $S_z (t)$ for different representative values of $F$ and for different structures and number of 
electrons $N_e$. This also gives us the opportunity to explore the dependence of the ultrafast demagnetization on 
band filling. Since the excitation spectrum depends on the precise structure and band filling of the 
model, and in order that the results can be compared, we have chosen the laser wave length such that it 
matches the 
absorption spectrum.
The results of Figs.~\ref{fig:figure1} and~\ref{fig:figure2} show that similar laser-induced 
demagnetizations 
take place for all considered geometries and band fillings. One observes that $S_z (t)$ decreases rapidly after the 
pulse passage at $t = 0$ ($\tau_p = 5$~fs) reaching values close to the long-time limit $S_z^\infty$ in $50$--$100$~fs. 
The characteristic demagnetization time scale $\tau_\text{dm}$ can be obtained by fitting an
exponential law of the form 
$S_z(t) = S_z^\infty + (S_z(0) - S_z^\infty) \exp \lbrace -t/\tau_\text{dm} \rbrace$ to the exact calculated numerical 
propagations, where $S_z(0)$ is the spin polarization at the time $t = 0$ when the electric field 
amplitude $E(t)$ reaches its maximum. The pulse shape is illustrated at the bottom panel of 
Fig.~\ref{fig:figure1}.\cite{footnote-fit} From the fits, shown as thin dashed curves in the 
figures, one obtains $\tau_\text{dm} \simeq 23$~fs for the rhombus and $\tau_\text{dm} 
\simeq 20$--$50$~fs 
for the triangles. These values, which correspond to $| \xi | = 80$~meV, are 3--5 times longer 
than the spin-orbit time 
scale $\hbar / \xi = 8$~fs. In Ref.~\onlinecite{Toe15} it has been shown that $\tau_\text{dm} \propto 
\hbar / \xi$ and that the rate of spin-to-orbital angular momentum transfer controls the UFD 
dynamics. 
Notice, moreover, that the spin relaxation occurs essentially after the passage of the laser pulse. 
In particular, 
$\tau_\text{dm}$ is always much larger than the considered pulse duration $\tau_p = 5$~fs. This implies that the 
demagnetization 
effect is not the direct result of the interaction with the laser electric field, but rather the 
consequence of an intrinsic process occurring within the 
excited electronic system. The same previous work shows that the interplay between the 
electronic motion in the 
lattice and the SOC is at the origin of the ultrafast demagnetization.\cite{Toe15} The spin-orbit 
interactions acting 
on the excited electrons bring about a local flow of angular momentum from the atomic spins $\vec s_i$ to the atomic 
orbits $\vec l_i$ on a time scale of the order of $\hbar / \xi = 8$~fs. At the same time, the 
hopping of the 
electrons between different atoms quenches any incipient increase of the total orbital angular 
momentum $\vec L = \sum_i \vec l_i$ on a very 
short time scale of only $\hbar / t_{jk}^{\alpha \beta} \lesssim 1$~fs. This prevents any 
accumulation of the 
transfered spin angular momentum in the orbital degrees of freedom. The local character of the above discussed 
mechanism of angular momentum transfer, including the laser excitation dominated by intra-atomic dipole 
transitions, supports the physical validity of the present small-cluster exact-propagation approach (see 
Sec.~\ref{sec:application}).

Figures~\ref{fig:figure1} and~\ref{fig:figure2} clearly show that the larger the fluence $F$ the larger the 
demagnetization $\Delta S_z = S_z^0 - S_z^\infty$, where $S_z^0$ denotes the initial spin polarization. For instance, in 
the triangle with $N_e = 7$ electrons the 
long-time spin polarization decreases 
from $S_z^\infty = 0.36 \, \hbar$ to $S_z^\infty = 0.09 \, \hbar$ per atom when the fluence is 
increased from $F = 10$ to $80~\text{mJ/cm}^2$. However, the characteristic 
shape of $S_z (t)$, and in 
particular the demagnetization 
time $\tau_\text{dm}$, depend weakly on $F$. To clarify this point, the insets in 
Figs.~\ref{fig:figure1} 
and~\ref{fig:figure2} show, on the right hand side, the scaled spin magnetization $S_\text{sc}(t) = 
[S_z(t) - S_z^\infty]/ \Delta S_z$ 
as a function of time $t$ for all considered fluences $F$. In addition, on the left hand side, the 
demagnetization time 
$\tau_\text{dm}$ is given as a function of $F$. One observes that for all 
considered systems $S_\text{sc}(t)$ 
and $\tau_\text{dm}$ are essentially independent of $F$, i.e., of the degree of excitation ($10~\text{mJ/cm}^2 \leq F 
\leq 80~\text{mJ/cm}^2$). However, $\tau_\text{dm}$ depends to some extent on 
the lattice structure and band filling, although it always remains in the range of a few tens of femtoseconds for 
$|\xi| = 80$~meV. This can 
be understood by 
recalling that the coupling between spin and translational degrees of freedom, which results from spin-orbit 
interactions, can be very sensitive to the details of 
the electronic 
structure. In fact, it is well-known that the magneto-crystalline anisotropy energy, 
easy magnetization axis, and 
orbital moments of transition-metal systems depend strongly on lattice structure and band 
filling.\cite{Pas95,Dor98,Rod98,Gui03} Furthermore, notice that weak 
oscillations are superimposed to the general exponential decrease of the calculated $S_z (t)$. These 
become somewhat weaker 
(stronger) 
for shorter (longer) pulse durations $\tau_p$, as the laser-field spectrum becomes broader (narrower) and the final 
excited state involves a larger (smaller) number of eigenfrequencies. They are possibly a 
consequence of the discreteness of the energy spectrum of the small cluster models used for the 
numerical 
time propagations.

It is instructive to compare our theoretical results for the fluence dependence of the UFD effect 
with available experiments.\cite{Koo10,Ten18} The measurements on Ni by Koopmans \textit{et 
al.}~have shown that with increasing fluence the relative demagnetization becomes 
stronger and the demagnetization time $\tau_\text{dm}$ increases.\cite{Koo10} While the former is in 
agreement with our trends, the latter is in clear contrast. However, more recent experiments on Ni 
by Tengdin \textit{et al.}~show a qualitatively different fluence dependence, namely, a 
fluence-independent demagnetization time\cite{Ten18}, which coincides with the predictions of our 
model (see Figs.~\ref{fig:figure1} and~\ref{fig:figure2}). Interestingly, Tengdin 
\textit{et al.}~fitted 
their experimental magnetization data by using up to three distinct 
exponential functions: The first one describes the initial laser-triggered magnetization decrease, 
which is investigated in this paper, whereas the 
remaining ones correspond to the subsequent magnetization recovery. In this way, they were 
able to separate the time scale of the ultrafast magnetization collapse from the much slower 
magnetization recovery. Concerning the latter process, both experimental groups have 
clearly observed that it depends qualitatively 
and quantitatively on the fluence of the 
pumping laser, and thus, on the absorbed energy. It is therefore possible that the fluence 
dependence 
of $S_z (t)$ observed in Ref.~\onlinecite{Koo10}, and the demagnetization time inferred using a 
single exponential fit, are partly affected by these energy dissipation processes, particularly as 
the level of excitation increases.

\begin{figure}[t]
 \centering
 \includegraphics[width=8.5cm,keepaspectratio=true]{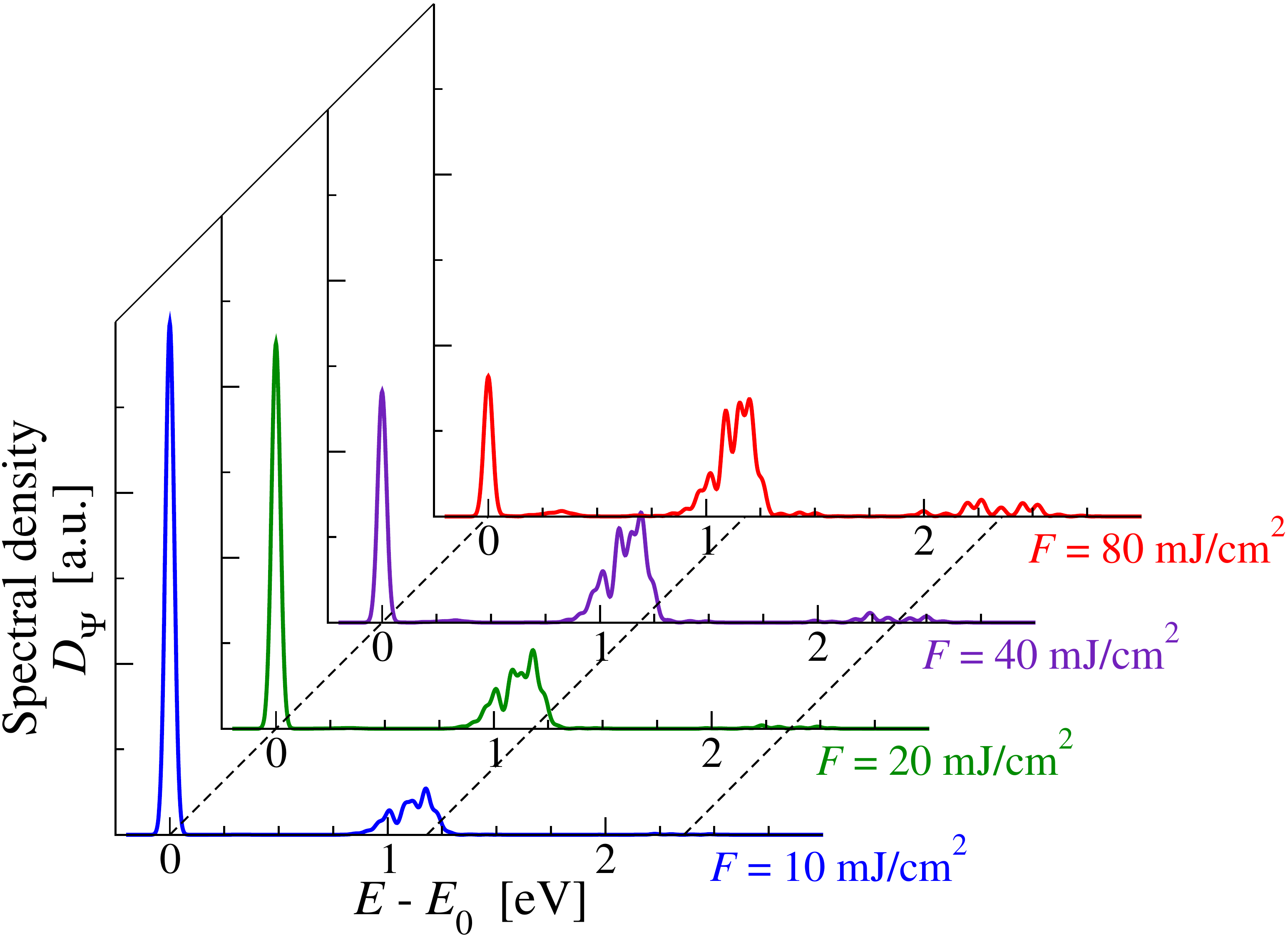} %width=8.663cm
 % figure1.eps: 0x0 pixel, 300dpi, 0.00x0.00 cm, bb=(atend)
 \caption{\small Spectral density $D_\Psi (\epsilon)$ of the excited many-body state $\ket{\Psi 
(t)}$ after the laser-pulse passage ($t \geq 15$~fs) as a function of excitation energy $\epsilon = E - E_0$. Results 
are given for the equilateral triangle having $N_e = 4$ and representative laser fluences $F$. The laser wave 
length is $\lambda = 1051$~nm ($\hbar \omega = 1.18$~eV). The discrete spectral lines have been broaden with a finite 
width $\delta = 20$~meV for the sake of clarity. The ground-state energy and the excitation energies corresponding to 
the absorption of one and two photons are indicated by the dashed lines.}
 \label{fig:figure3}
\end{figure}

It is interesting to investigate the experimentally observed increase of the demagnetization $\Delta 
S_z$ with increasing fluence $F$
by analyzing the spectral distribution of the many-body state 
$\ket{\Psi (t)}$ after the pump-pulse passage (e.g., $t 
\geq 15$~fs for a $5$~fs laser pulse). For this purpose, we expand $\ket{\Psi} = \sum_k \alpha_k \, \ket{\psi_k}$ in 
the stationary states $\ket{\psi_k}$ of the field-free Hamiltonian $\hat H = \hat H_0 + \hat H_C + \hat H_\text{SO}$ 
satisfying $\hat H \ket{\psi_k} = E_k \ket{\psi_k}$. The spectral distribution of $\ket{\Psi}$ is then given by
\begin{equation}
  D_{\Psi} (\epsilon) = \sum_{k} | \skp{\psi_k}{\Psi (t)} |^2 \, \delta(\epsilon - \epsilon_k)~,
\end{equation}
where $\epsilon = E - E_0$ is referred to the ground-state energy $E_0$ and $\epsilon_k = E_k - 
E_0$. Notice that $\hat H$ and thus the spectral distribution $D_{\Psi} (\epsilon)$ of $\ket{\Psi(t)}$ are independent 
of $t$ once the pulse has passed (e.g., $t \geq 3 \tau_p$). Fig.~\ref{fig:figure3} shows $D_{\Psi} (\epsilon)$ for a 
triangle with $N_e = 4$, which has been excited with a $5$~fs laser pulse of wave length $\lambda = 1051$~nm ($\hbar 
\omega = 1.18$~eV) and fluences $F = 10$, $20$, $40$ and $80 ~ \text{mJ/cm}^2$. Three main peaks or groups of nearby 
peaks are distinguished around $\epsilon = 0$, $\hbar \omega$ and $2 \hbar \omega$. They correspond to the ground state 
and to the absorption of $1$ and $2$ photons. One observes how the spectral weight of the 
excited-state manifolds 
around $\hbar \omega$ and $2 \hbar \omega$ increases with increasing $F$ at the expense of the ground-state 
contribution $|\skp{\Psi_0}{\Psi(t)}|^2$. This reflects the growing level of electronic excitation and can be directly 
related to the degree of demagnetization $\Delta S_z / S_z^0 = (S_z^0 - S_z^\infty)/S_z^0$ achieved at long times $t 
\gg \tau_\text{dm}$. Indeed, a simple argument allows us to approximately express $\Delta S_z / 
S_z^0$ in terms of the angle $\alpha = \arccos \, \skp{\Psi_0}{\Psi(t)}$ between the excited state 
$\ket{\Psi(t)}$ at $t \geq 3 \tau_p$ 
and 
the ground state $\ket{\Psi_0}$. Writing
\begin{equation}\label{eq:alpha1}
  \ket{\Psi(t)} = \cos (\alpha) \, \ket{\Psi_0} + \sin(\alpha) \, \ket{\Delta \Psi (t)} 
\end{equation}
with $\skp{\Psi_0}{\Delta \Psi(t)} = 0$ and $\skp{\Delta \Psi(t)}{\Delta \Psi(t)} = 1$ we have
\begin{equation}\label{eq:alpha2}
  S_z (t) = \cos^2 (\alpha) \, S_z^0 + \sin^2 (\alpha) \, \bra{\Delta \Psi (t)} \hat S_z \ket{\Delta \Psi(t)} ~.
\end{equation}
The demagnetization in the long-time limit is then given by
\begin{equation}\label{eq:alpha3}
  \Delta S_z = \sin^2 (\alpha) \left[ S_z^0 - S_{z}^* (\infty) \right] ~,
\end{equation}
where $S_{z}^* (t) = \bra{\Delta \Psi(t)} \hat S_z \ket{\Delta \Psi(t)}$ is the magnetization in the excited states at 
time $t$. This shows that $\Delta S_z / S_z^0$ is proportional to the spectral weight $\sin^2 (\alpha)$ transfered to 
the excited states or, in other words, to the level of excitation. Since $S_z^* (0) \simeq S_z^0$ 
(i.e., essentially no 
change in the spin polarization occurs during the pulse passage) the proportionality factor $S_z^0 - 
S_z^* (\infty) = S_z^* (0) 
- S_z^*(\infty)$ gives a measure of the efficiency of the demagnetization in the excited-state 
manifolds. It is interesting to observe that the dynamics of the many-electron system yields a 
remarkably 
effective reduction of the excited-state magnetization $S_{z}^* (t)$. In fact, in some cases (e.g., 
a triangle having $N_e = 7$ electrons) the 
quenching of $S_{z}^* (t)$ is nearly complete (i.e., $S_z^* (\infty) \simeq 0$). While it is 
tempting to 
interpret this in terms of the statistical hypothesis of equal a priori probability, there are many 
examples where no full excited-state quenching is found. For instance, in a triangle 
having $N_e = 4$ or $5$ electrons, as well as the rhombus, one finds that $S_z^* (\infty)$ is 
significantly larger than zero ($S_z^* (\infty) \simeq 0.06$--$0.13 \, \hbar$ per atom, see 
Sec.~\ref{subsec:lev-exc}).

At this stage one may wonder whether the relation between the degree of long-time demagnetization 
and the level of 
excitation is not simply a consequence of the fact that with increasing fluence $F$ and increasing $\sin^2 (\alpha)$ 
also 
the absorbed energy $\Delta E$ increases. In order to clarify this matter it is important to 
investigate the dynamical magnetic response as a function of the 
photon energy $\hbar \omega$.

\subsection{Absorbed energy versus average number of absorbed photons}
\label{subsec:lev-exc}

The preceding section has shown that the main consequence of increasing the level of 
electronic 
excitation is to enhance the degree of demagnetization $\Delta S_z = S_z^0 - S_z^\infty$ at long 
times, at least for the considered range of fluence $F$.
A complementary way of investigating the dependence of ultrafast demagnetization 
on the level of excitation and on the absorbed energy $\Delta E$ is 
to vary systematically the photon 
energy $\hbar \omega$. In this way the importance of the absorbed energy and of the average number 
of electrons excited by the laser or of absorbed photons $n_\text{ph} = \Delta E / \hbar \omega$
can be tell apart.

In the following different laser frequencies are considered, for which the absorption probabilities 
are significant. The corresponding exact time dependences of $\ket{\Psi(t)}$ and $S_z(t)$ have been 
numerically determined. In all cases, the UFD effect is observed 
with demagnetization times $\tau_\text{dm} = 18$--$62$~fs for the triangle with $N_e = 4$ electrons, $\tau_\text{dm} = 
23$--$62$~fs for the triangle with $N_e = 
5$, $\tau_\text{dm} = 42$--$122$~fs for the triangle with $N_e = 7$, and $\tau_\text{dm} = 15$--$86$~fs for the 
rhombus with $N_e = 5$. This confirms that the UFD effect is an intrinsic 
characteristic of the 
correlated electronic system, which is qualitatively independent of the details of the 
triggering excitation.
Nevertheless, notice that the precise value of $\tau_\text{dm}$ depends to some extent on the laser 
frequency $\omega$. This shows that different optical absorptions lead to different excited states, 
or more generally, different spectral distributions $D_\Psi (\epsilon)$, which exhibit their own 
specific many-body dynamics. Incidentally, this may also indirectly cause a fluence dependence of 
$\tau_\text{dm}$. Assuming a rapid thermalization of the electronic translational degrees of 
freedom after the laser absorption, one expects that the distribution of the excited many-body 
states should become broader as the fluence $F$ increases. This would render higher excitation 
energies accessible and could thus result in changes in $\tau_\text{dm}$ as a function of $F$. 
Unfortunately, this hypothesis cannot be quantified numerically in the present framework, since the 
cluster models accessible to exact time propagations are too small to allow a true thermalization or 
self-averaging.\cite{Deu91}

\begin{figure}[b]
 \centering
 \includegraphics[width=7.5cm,keepaspectratio=true]{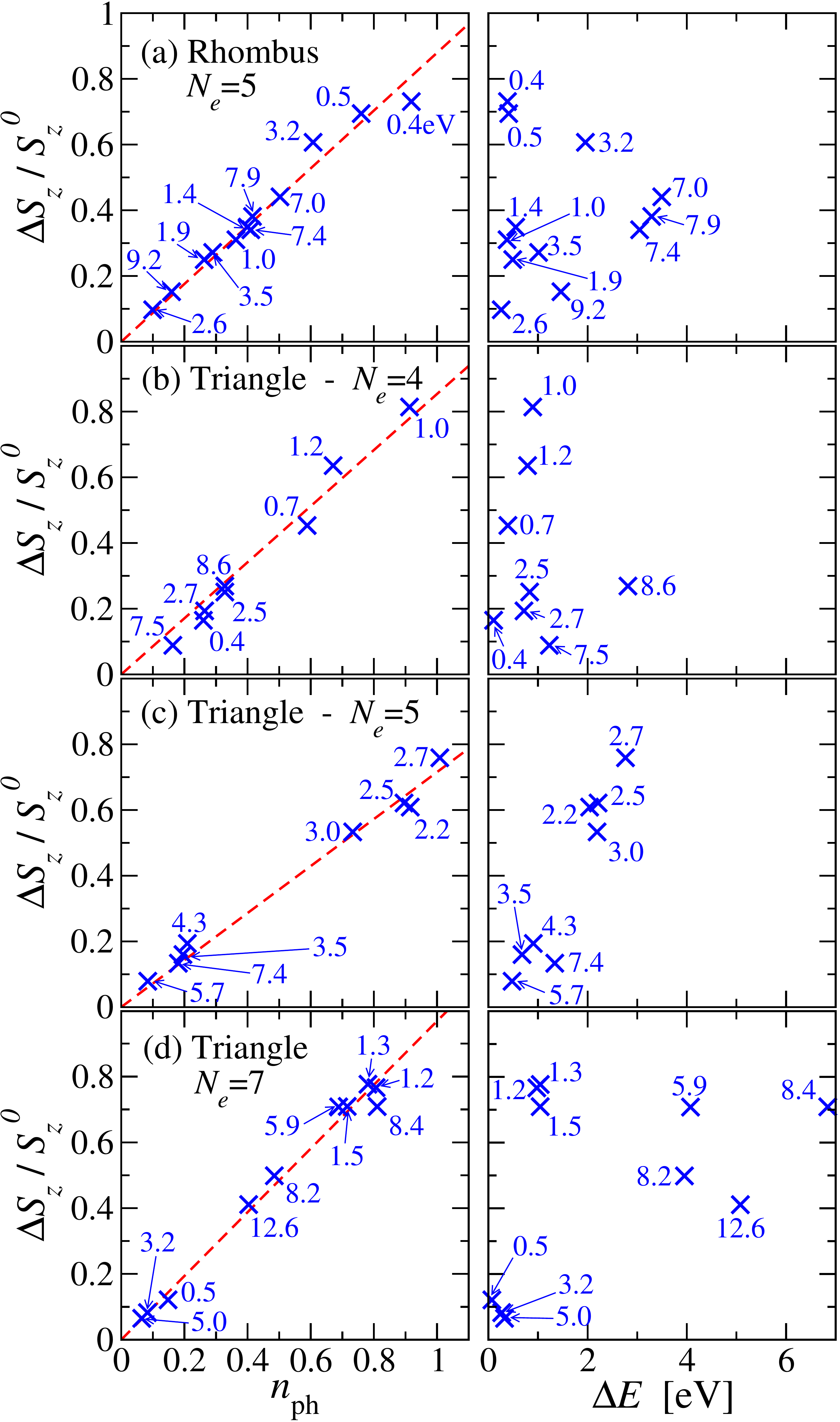} %width=8.663cm
 % figure1.eps: 0x0 pixel, 300dpi, 0.00x0.00 cm, bb=(atend)
 \caption{\small Correlation between the laser-induced degree of demagnetization $\Delta S_z / 
S_z^0$ at long times $t \gg \tau_\text{dm}$ and the absorbed energy $\Delta E$ (right) or
the average number of absorbed photons $n_\text{ph} = \Delta E / \hbar \omega$ (left), where 
$\omega$ 
is the angular 
frequency of the exciting laser. The crosses are obtained from the exact calculated time evolution 
for (a)~a rhombus 
with $N_e = 5$ electrons, and an equilateral triangle having (b)~$N_e = 4$, (c)~$N_e = 5$ and 
(d)~$N_e = 
7$ electrons. The laser fluence is always $F = 40~\text{mJ/cm}^2$, while the different photon 
energies $\hbar \omega$ 
are indicated in eV. The dashed straight lines on the left panels are fits to the approximate linear 
dependence of 
$\Delta S_z / S_z^0$ 
on $n_\text{ph}$.}
 \label{fig:figure4}
\end{figure}

The long-time limit of the demagnetization $\Delta S_z = S_z^0 - S_z^\infty$ has been 
derived for each $\hbar \omega$ from the numerical time propagations. The thus obtained relative demagnetizations 
$\Delta S_z / S_z^0$ are shown in Fig.~\ref{fig:figure4} as a function of the absorbed energy $\Delta E$ 
(right-hand side) and of the average number of absorbed photons $n_\text{ph} = \Delta E / \hbar 
\omega$ (left-hand side). 
The scatter plot on the right-hand side of Fig.~\ref{fig:figure4} is so disperse that no relation 
between 
$\Delta E$ and $\Delta S_z / S_z^0$ can be established. $\Delta S_z$ is obviously not a 
function of $\Delta E$ alone. In contrast, the left-hand side figure reveals a remarkably simple, 
approximately linear dependence of $\Delta S_z / S_z^0$ on $n_\text{ph}$.
For example, in the 
triangle with $N_e=7$, $\Delta S_z / S_z^0$ are nearly the same for $\hbar \omega = 1.5$~eV 
and $\hbar \omega = 5.9$~eV although the absorbed energies $\Delta E = 1.0$~eV and $\Delta E = 4.1$~eV  
differ widely by a factor four. The corresponding $n_\text{ph} = 0.72$ and $n_\text{ph} = 0.69$ are very similar. In 
other cases, for example in the triangle with $N_e=4$, the absorbed energies are very similar 
($\Delta E = 0.9$~eV for 
$\hbar \omega = 1.0$~eV and $\Delta E = 0.8$~eV for $\hbar \omega = 2.5$~eV) but the relative demagnetizations 
differ widely ($\Delta S_z / S_z^0 = 0.81$ and $\Delta S_z / S_z^0 = 0.25$, respectively). One 
concludes that the average 
number of absorbed photons $n_\text{ph}$, or equivalently, the number of single-particle electronic 
excitations induced by the pumping pulse, rather than the absorbed energy, determines 
primarily the strength of the 
demagnetization. This is consistent with the discussion at the end of Sec.~\ref{subsec:fluence} showing that $\Delta S_z 
/ S_z^0$ is proportional to the spectral weight transfered to the excited
states during the laser-pulse absorption.

The slope $\gamma$ of the linear dependence $\Delta S_z / S_z^0 \simeq \gamma \, n_\text{ph}$ can 
be related to the 
efficiency of the demagnetization in the excited states $S_z^*(\infty)$, which was introduced at the end of 
the Sec.~\ref{subsec:fluence}. Assuming for simplicity that only the ground state and the lowest excited 
states around $\hbar \omega$ contribute to the spectral distribution of $\ket{\Psi(t)}$ after the 
pump pulse, one can easily show that $\sin^2 (\alpha)$ in Eqs.~\eqref{eq:alpha1}--\eqref{eq:alpha3} 
is equal to the average number of 
absorbed photons $n_\text{ph} = \Delta E / \hbar \omega$. Therefore, $\gamma = 
\left[ S_z^0 - S_z^*(\infty) \right] / S_z^0$ represents the relative demagnetization efficiency in 
the 
excited states. Figure~\ref{fig:figure4} (left) shows that $\gamma$ and thus $S_z^*(\infty)$ do 
not depend significantly on $\hbar \omega$. However, they depend somewhat on the 
considered cluster model and band filling. For example, for the triangle with $N_e = 4$ and $5$ electrons, and for the 
rhombus with $N_e = 5$ electrons we find $\gamma \simeq 0.72$--$0.88 < 1$. This implies that the magnetic order in 
the excited states $\ket{\Delta \Psi}$ 
is not fully destroyed as a result of the many-electron dynamics. In other words, $\ket{\Delta 
\Psi}$ remains 
ferromagnetic to a small extent even at very long times.
In contrast, for the triangle with $N_e = 7$, the FM correlations in $\ket{\Delta \Psi}$
are fully lost along the dynamics.
The demagnetization of $\ket{\Delta \Psi}$ is in this case almost complete, namely, $S_z^* 
(\infty) / S_z^0 \simeq 0.03$ or $\gamma \simeq 0.97$.

According to our exact model calculations, the energy per atom $\Delta E$, which is absorbed 
during the pump pulse, does not give the appropriate measure of the degree of excitation of the 
electronic 
system in relation to subsequent $\Delta S_z / S_z^0$. This is physically interesting, since it 
contrasts with the idea that the translational degrees of freedom of the electronic system should 
rapidly thermalize in a spin-conserving way. Indeed, if the latter were so, the energy absorbed in 
any field-induced single-particle transition would be rapidly redistributed among the electrons, 
thus erasing any memory of the details of the triggering excitation (e.g., the number of initial 
single-particle transitions or number of absorbed photons). Let us recall that the characteristic 
times involved in electron-lattice and electron-electron interactions ($\hbar / 
t_{ij}^{\alpha\beta}$ and $\hbar / U$) are at least an order of magnitude shorter than the typical 
spin-orbit and demagnetization times. All these short-time dynamical processes are properly taken 
into account in our studies. Still, it is also true that our calculations are unable to describe the 
approach to thermal equilibrium, since the exact time propagations are performed for closed purely 
electronic systems (Neumann-Liouville theorem). The interactions with the 
environment are ignored and the considered models are too small to achieve 
self-averaging.\cite{Deu91} It is 
unclear at present what would be the characteristic time involved in the thermalization of the 
translational electronic degrees of freedom of ferromagnetic metals, and 
how such a thermalization would affect the relation between $\Delta S_z / S_z^0$ and the absorbed 
energy $\Delta E$. Extensions of our calculations by taking into account a spin-conserving coupling 
to a bath, which simulates the environment, as well as numerical time propagations of mixed states 
corresponding to translationally thermalized electronic states are therefore worthwhile.

\subsection{Electric-field polarization}
\label{subsec:polarization}

\begin{figure}[t]
 \centering
 \includegraphics[width=8.5cm,keepaspectratio=true]{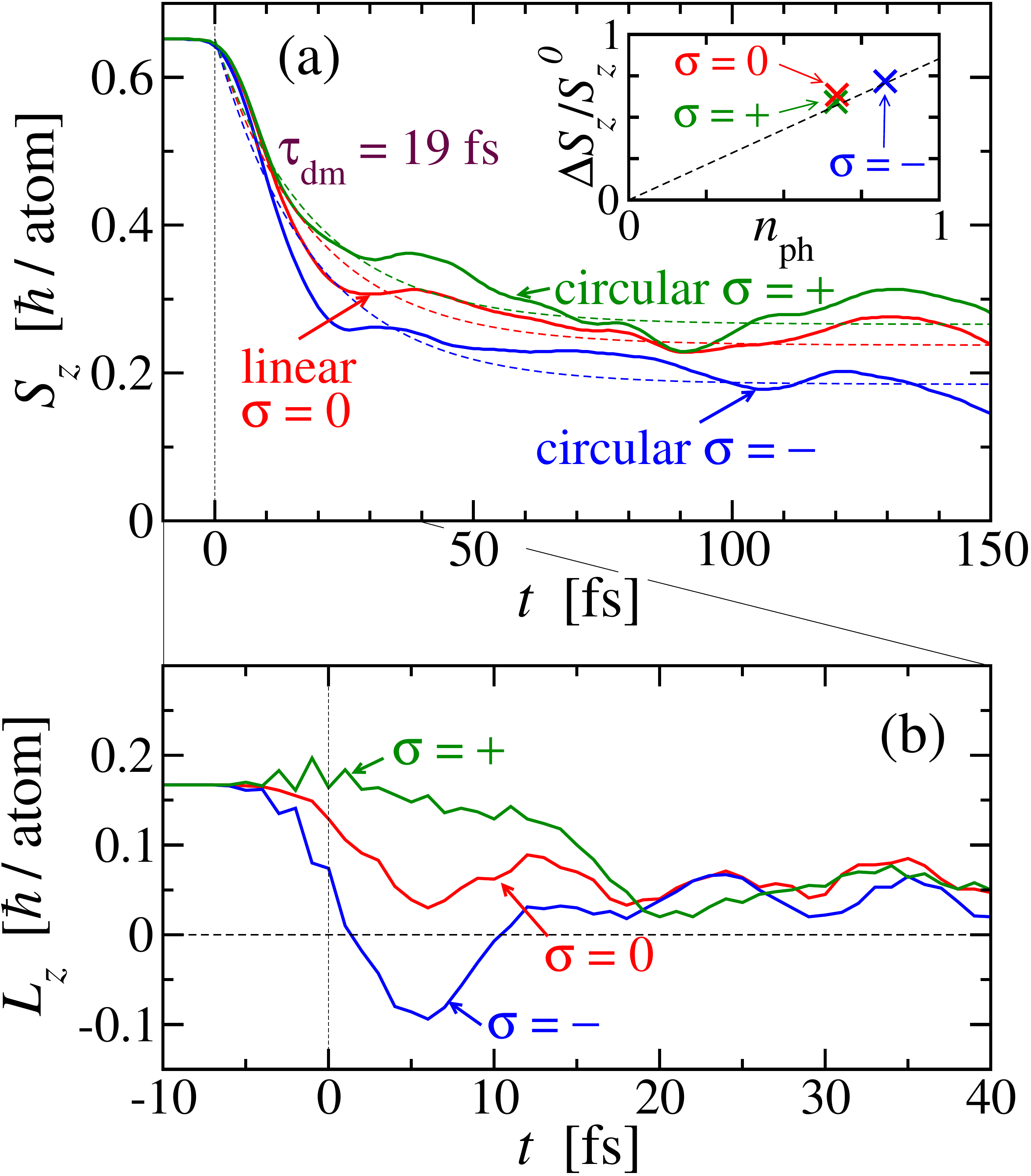} %width=8.663cm
 % figure1.eps: 0x0 pixel, 300dpi, 0.00x0.00 cm, bb=(atend)
 \caption{\small Time dependence of the average (a)~spin moment $S_z$ and (b)~orbital moment $L_z$ following 
the excitation with a $\tau_p = 5$~fs laser pulse having a wave length $\lambda = 1051$~nm and a 
polarization $\hat \varepsilon$ which is in-plane linear ($\sigma = 0$), right circular ($\sigma = 
+$) or left circular ($\sigma = -$). 
The full curves are obtained from the exact time evolution of the equilateral triangle model with $N_e = 4$ 
electrons. The dashed curves in~(a) are exponential fits to $S_z (t)$ with a common demagnetization time 
$\tau_\text{dm} = 19$~fs. The inset in~(a) shows the degree of 
demagnetization $\Delta S_z / S_z^0$ as a function of the average number of absorbed 
photons $n_\text{ph} = \Delta E / \hbar \omega$, together with the same linear approximation (dashed line) as the one 
found in Fig.~\ref{fig:figure4}(b) left.}
 \label{fig:figure5}
\end{figure}

The dependence of the magnetization dynamics on the polarization $\hat \varepsilon$ of the incident 
laser 
pulse has been investigated by considering linearly 
and circularly polarized electric fields. Figure~\ref{fig:figure5} shows the time dependence of 
the spin and 
orbital angular momenta in an equilateral triangle with $N_e = 4$ electrons. The pumping
excitation has a 
duration $\tau_p = 5$~fs and a wave length $\lambda = 1051$~nm. Three different electric-field polarizations 
$\hat \varepsilon$ are considered: linear polarization along a NN bond within the $xy$-plane 
containing the 
triangle ($\sigma = 0$), right circular polarization ($\sigma = +$) and left circular polarization ($\sigma = -$). For 
$\sigma = +$ ($\sigma = -$) the field carries an angular momentum of $\hbar$ ($-\hbar$) which is parallel 
(antiparallel) to the ground-state spin magnetization $S_z^0$ along the out-of-plane $z$ 
direction. Figure~\ref{fig:figure5}(a) shows that $S_z (t)$ depends weakly on the 
considered 
polarization, in agreement with experiment.\cite{Dal07} In the limit of long times, 
the spin magnetization decreases to a somewhat larger (smaller) value $S_z^\infty$ after the 
absorption of a right (left) 
circular pulse in comparison with the linear pulse. As we shall see, this 
can be ascribed to the rather small polarization dependence of the absorption cross section. One may 
also notice that the difference 
in $S_z (t)$ between left and right polarized light increases at the early stages of the dynamics ($t \lesssim 
\tau_\text{dm} = 19$~fs) showing some oscillations for $t \geq 
\tau_\text{dm}$.\cite{footnote-polarizations}

The laser-polarization effects on the orbital magnetic moment $L_z$ are found to be significant only 
at very short times 
($t \lesssim 10$--$15$~fs). The dynamics of the initially quenched moment $L_z \simeq 0.17 \, \hbar / \text{atom}$, 
which is parallel to $S_z$, is shown in Fig.~\ref{fig:figure5}(b) for different laser polarizations. 
For linear polarization ($\sigma = 0$), the orbital moment decreases to around $L_z \simeq 
0.03 \,\hbar / 
\text{atom}$ during the action of the pulse, while in the case of a left (right) circularly polarized pulse $L_z$ 
decreases by around $0.26 \,\hbar / \text{atom}$ (increases by around $0.03 \,\hbar / \text{atom}$) on the same time 
scale. Notice that for $\sigma = -$, $L_z$ becomes even negative (i.e., antiparallel to $S_z$) for a very short time.
The time lapse during which the polarization dependence of $L_z$ is significant is of the order of 
the pulse width, in 
the present case $\tau_p = 5$~fs.

In order to analyze the polarization effects on $L_z$, let us first 
notice that 
the absorption of non-polarized $\sigma = 0$ pulses consists in electronic dipole transitions mainly from the $3d, 
m=\pm 1$ orbitals to the $4p, m=0$ orbitals. Thus, the orbital-moment projection is reduced, which 
explains qualitatively the 
laser-induced decrease of $L_z$ observed for linearly polarized pulses [see Fig.~\ref{fig:figure5}(b)].
For $\sigma = -$ ($\sigma = +$) $L_z$ is further decreased (enhanced) by around $0.12$--$0.17 \, \hbar / 
\text{atom}$ in comparison with the $\sigma = 0$ dynamics ($t \lesssim 
10$--$15$~fs). This 
polarization-dependent decrease (enhancement) is the consequence of 
the transfer of angular momentum from the laser field to the orbital electronic motion. The left (right) polarized 
light induces $m \to m-1$ ($m \to m+1$) intra-atomic transitions, where the azimuthal quantum number $m$ gives the 
local contribution to $L_z$. Neglecting for a moment any spin-orbit transitions and interatomic electron hoppings, 
this would imply a change $\Delta L_z = \pm \hbar$ in the angular momentum per absorbed photon. Knowing that 
$n_\text{ph} / N_a \cong 0.28$ ($n_\text{ph} / N_a \cong 0.22$) for left (right)
polarization, we conclude that the change in $L_z$ induced per absorbed photon explains qualitatively the observed 
short-time decrease (enhancement) of $L_z$ for $\sigma = -$ ($\sigma = +$). The 
polarization-dependent change $\Delta 
L_z$ is in fact somewhat smaller than $\pm n_\text{ph} \hbar / N_a$, since part of the effect is lost due to the rapid 
interatomic hoppings.

It is also important to remark that the changes in $L_z$ induced by the laser field, and the thus resulting differences 
in the time dependence of $L_z$ for different polarizations, rapidly vanish once the laser pulse passes. 
As shown in Fig.~\ref{fig:figure5}(b), already $18$~fs after the pulse reaches its maximum ($t=0$) the differences 
in $L_z (t)$ for different $\sigma$ are no longer distinguishable from the intrinsic oscillation of $L_z (t)$ due to 
the dynamics ruled by the field-free $\hat H$. The reason behind this is the motion of $3d$ electrons throughout the 
lattice, which does not conserve the atomic $l_{iz}$. In TMs $d$-electron delocalization actually 
quenches $L_z$ on 
a very short time scale of the order of $\hbar / t_{ij}^{\alpha\beta} \simeq 1$~fs, where 
$t_{ij}^{\alpha\beta}$ is the 
hopping integral between NNs. Thus, the electronic motion tends to wash out any change in the 
orbital angular momentum, irrespectively of its origin. The results show that the hopping-induced 
rapid quenching of $L_z$ applies 
equally well to an enhancement of $L_z$ due to the laser absorption ($\sigma = +$) and to the 
spin-to-orbital 
angular momentum transfer due to SOC in the excited states. This explains why the time dependences of $L_z (t)$ for the 
different laser polarizations are very similar after the pulse passage. The differences in the 
excited state for different $\sigma$, which are clearly visible in $L_z (t)$ for 
short times, have only a modest effect on the slower spin dynamics [see Fig.~\ref{fig:figure5}(a)]. 
The 
latter is actually governed by the spin-to-orbital transfer of angular momentum and the 
above-mentioned
$L$-quenching electronic motion. As we shall see, the dependence of $S_z (t)$ on the laser polarization is mainly due 
to the changes in the absorption efficiency for different $\sigma$. One concludes that the $pd$ model explains from 
a microscopic perspective the experimentally observed weak sensitivity of the UFD effect on the laser 
polarization.\cite{Dal07}

In the present calculations the same fluence $F = 40~\text{mJ/cm}^2$ has been 
used for all electric-field polarizations. The obtained degrees of excitation, as measured by $n_\text{ph} = \Delta E / 
\hbar \omega$ and the long-time 
demagnetization $\Delta S_z / S_z^0$, are quantitatively similar for all $\sigma$ [see the inset in 
Fig.~\ref{fig:figure5}(a)]. 
Furthermore, Fig.~\ref{fig:figure5}(a) shows 
that the time dependences of $S_z(t)$ for different $\sigma$ can all be reasonably well fitted with 
exponential functions having the same demagnetization time $\tau_\text{dm} = 19$~fs (dashed curves). 
Our calculations show no significant effect 
of the laser polarization $\hat \varepsilon$ on $\tau_\text{dm}$.

\begin{figure}[t]
 \centering
 \includegraphics[width=8.5cm,keepaspectratio=true]{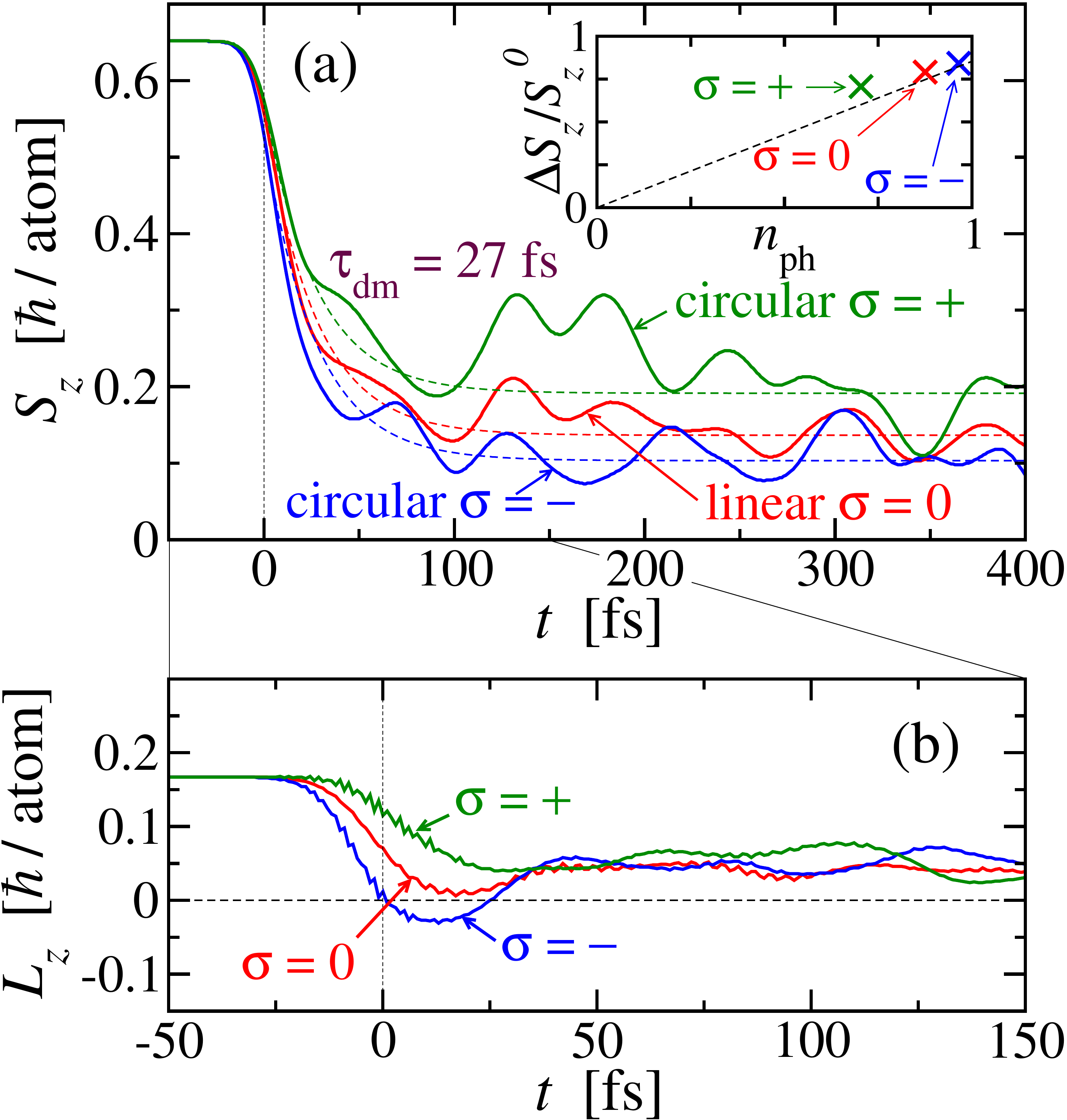} %width=8.663cm
 % figure1.eps: 0x0 pixel, 300dpi, 0.00x0.00 cm, bb=(atend)
 \caption{\small Time dependence of the average (a)~spin moment $S_z$ and (b)~orbital moment $L_z$ 
corresponding to an excitation with a $\tau_p = 20$~fs laser pulse, having a wave length $\lambda = 
1051$~nm and a polarization $\hat 
\varepsilon$ which is in-plane linear ($\sigma = 0$), right circular ($\sigma = +$) and left circular ($\sigma = -$). 
See also the caption of Fig.~\ref{fig:figure5}.}
 \label{fig:figure6}
\end{figure}

In order to investigate the interplay between spin-orbit coupling and laser-ferromagnet interaction, 
it is interesting to consider pulse durations $\tau_p$ that are larger than the time scale of the 
SOC ($\hbar / |\xi| = 8$~fs) for different laser polarizations. Figure~\ref{fig:figure6} shows 
the time dependences of $S_z$ and $L_z$ in an 
equilateral triangle ($N_e = 4$ electrons) which is excited with a laser pulse having $\tau_p = 
20$~fs and $\lambda = 1051$~nm. One observes that $S_ z (t)$ and $L_z (t)$ depend significantly on 
the considered polarization. In the case of $L_z$ the polarization-dependent changes resulting from 
direct optical absorption vanish very rapidly as the pulse passes ($t > \tau_p = 20$~fs). As already 
discussed, this is due to the rapid electron delocalization in the 
lattice [see Fig.~\ref{fig:figure6}(b)]. In contrast, the differences in $S_z (t)$ for the different
considered $\hat \varepsilon$ remain significant during several hundreds of femtoseconds [see 
Fig.~\ref{fig:figure6}(a)], well beyond the point where the electric field has vanished. The 
results also show that at long times the circular $\sigma = -$ ($\sigma = +$) pulse induces 
a more (less) efficient demagnetization $\Delta S_z / S_z^0$ than the linear $\sigma = 
0$ pulse. The actual values of $\Delta S_z / S_z^0$ for different $\hat \varepsilon$ correlate well 
with the average number of absorbed photons, as shown in the inset of Fig.~\ref{fig:figure6}(a). As 
for shorter pulses, the demagnetization time $\tau_\text{dm} = 27$~fs is found to be essentially 
independent of $\hat \varepsilon$.

The small polarization dependence of the long-time demagnetization degree $\Delta S_z 
/ S_z^0$ can be interpreted qualitatively in terms of the orbital occupations. Let us first recall 
that the initial state before the pulse absorption has a small positive orbital moment 
$L_z \simeq 0.17 \, \hbar$ per atom. This means that the $3d$ orbitals with $m > 0$ are in average more likely occupied 
than the orbitals with $m < 0$. This introduces an asymmetry in the absorption of left and right polarized light 
(dichroism). Since the orbital polarization of the final $4p$ states is negligible, and the optical 
matrix elements are 
invariant upon reversing the circular polarization and the sign of the initial-state $m$, a higher laser absorption is 
expected when the average occupation of the dominant initial states is larger. In the case of left (right) circularly 
polarized light the $m \to m-1$ ($m \to m+1$) selection rule implies that the absorption is dominated by the initial 
states having $m > 0$ ($m < 0$). Consequently, for $L_z > 0$ the absorption cross section for left-circularly polarized 
light should be somewhat larger. Our results confirm this trend and can be interpreted accordingly. 
For example, for a
$\tau_p = 5$~fs laser pulse we obtain that the average number of absorbed photons is $n_\text{ph} = 0.83$ for 
left-circularly polarized pulses, while it is about $n_\text{ph} = 0.67$ for linearly or right-circularly polarized 
pulses. 
Similarly, for a $\tau_p = 20$~fs laser pulse we obtain $n_\text{ph} = 0.96$ for left 
polarization, $n_\text{ph} = 0.88$ for linear polarization, and $n_\text{ph} = 0.70$ for right 
polarization. The insets in Figs.~\ref{fig:figure5} and~\ref{fig:figure6} show the already discussed linear dependence
between $\Delta S_z / S_z^0$ and $n_\text{ph}$. In fact, the slopes of the straight dashed lines in 
the insets of Figs.~\ref{fig:figure5}(a) and~\ref{fig:figure6}(a) are the same as in 
Fig.~\ref{fig:figure4}(b). One concludes that the 
dependence of $\Delta S_z / S_z^0$ on the laser polarization is mainly a consequence of the 
different absorption cross sections. It should be, however, noted that the orbital magnetic 
moments $\langle L_z \rangle$ in TMs are weak. In other words, the differences in the ground-state 
occupations for positive and negative $m$ are small. Therefore, the possibilities 
of taking advantage of dichroism in order to tune the degree of excitation and $\Delta S_z / S_z^0$ seem quantitatively 
limited.

\subsection{Pulse duration}
\label{subsec:duration}

\begin{figure}[t]
 \centering
 \includegraphics[width=7.5cm,keepaspectratio=true]{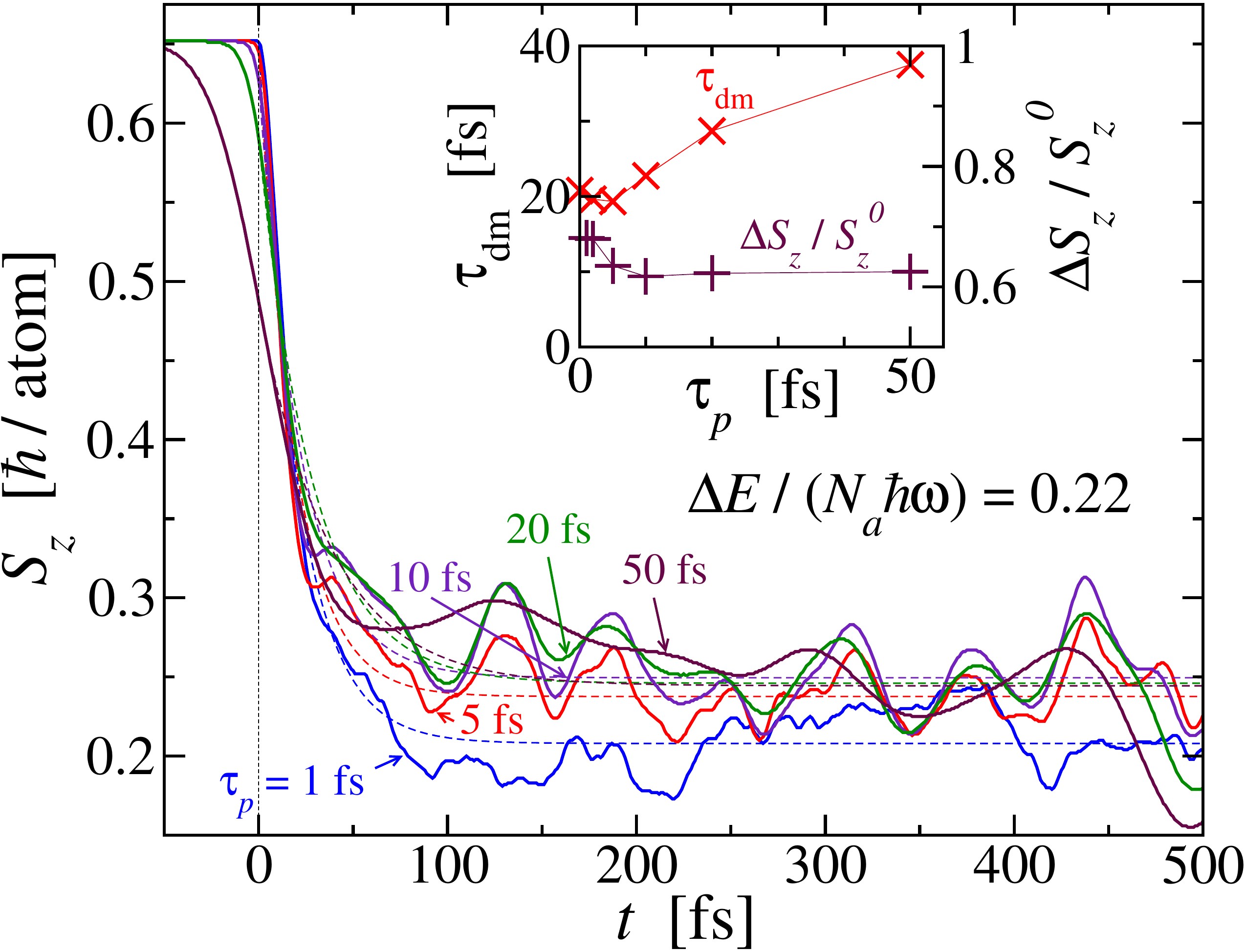} %width=8.663cm
 % figure1.eps: 0x0 pixel, 300dpi, 0.00x0.00 cm, bb=(atend)
 \caption{\small Time dependence of the average spin moment $S_z$ in an equilateral triangle with $N_e = 4$ 
electrons resulting from laser-pulse excitations having wave length $\lambda = 1051$~nm and pulse 
durations $\tau_p = 1$--$50$~fs. 
The full curves are obtained from the exact time evolutions, while the dashed curves are the 
corresponding exponential fits to $S_z (t)$. The laser fluences are such 
that the average number of absorbed photons per atom is $n_\text{ph} / N_a = \Delta E / (N_a 
\hbar \omega) = 0.22$ for all $\tau_p$. The inset shows the demagnetization time 
$\tau_\text{dm}$ and the long-time 
demagnetization $\Delta S_z / S_z^0$ as a function of $\tau_p$.}
 \label{fig:figure7}
\end{figure}

The pulse duration $\tau_p$ is a central characteristic of the laser excitation whose role on the dynamics deserves 
to be investigated in some detail. To this aim, exact time propagations have been performed for a 
triangle having $N_e = 4$ electrons, which is excited with a laser having $\lambda = 1051$~nm and 
$1~\text{fs} \, \leq \tau_p \leq \, 
50~\text{fs}$. This covers the range from narrow to broad pulses in comparison with the period of 
oscillation of the field $T = \lambda / c \simeq 3.5$~fs and the SO time scale $\hbar / |\xi| 
\simeq 8$~fs. Since the radiated energy is directly proportional to the 
pulse duration $\tau_p$, and the absorption efficiency depends strongly on the frequency distribution of the field, 
comparing the magnetization dynamics for the same fluence $F$ and different $\tau_p$ would be confusing. We have 
therefore scaled $F$ for each $\tau_p$ so that the absorbed energy $\Delta E$ and 
the average number of absorbed photons $n_\text{ph} = \Delta E / \hbar \omega$ remain constant. In this way the role 
of the pulse duration can be effectively assessed.
Figure~\ref{fig:figure7} shows the time dependence of the average spin magnetization $S_z (t)$ 
for $\tau_p = 1$--$50$~fs and $F$ such that $n_\text{ph} / N_a = 0.22$. For relatively 
short pulses ($\tau_p < 
10$~fs) the decrease of $S_z(t)$ takes place sharply after the pulse passage. The excitation is sudden, since the 
electronic system has no time to evolve from a magnetic point of view ($\tau_\text{dm} \simeq 20$~fs and $\hbar/|\xi| 
= 8$~fs for $|\xi| = 80$~meV). However, as the pulse duration is increased, one observes that a 
significant part of the demagnetization occurs while the laser field is still on. This is 
particularly clear for 
$\tau_p = 50$~fs, in which case almost half of the long-time demagnetization has already 
taken place when the 
laser pulse reaches its maximum at $t = 0$ (see Fig.~\ref{fig:figure7}).

The demagnetization time $\tau_\text{dm}$ and the degree of 
demagnetization $\Delta S_z / S_z^0$ are obtained by fitting the exact calculated time dependences 
of $S_z$ (full curves) for each $\tau_p$ with simple exponential functions. The 
results, given in the inset of Fig.~\ref{fig:figure7}, show 
that $\Delta S_z / S_z^0 = 0.62$--$0.68$ is essentially independent of the 
pulse duration, provided that $n_\text{ph}$ is kept constant. This holds even for pulse durations 
which are much longer than the spin-orbit time 
scale ($\tau_p > \hbar / |\xi| = 8$~fs), which is consistent with the fact that 
the long-time 
demagnetization $\Delta S_z / S_z^0$ is controlled primarily by the number of laser-induced 
single-particle excitations or absorbed photons (see 
Sec.~\ref{subsec:lev-exc}). At least in these examples, the simultaneous action of the laser field and 
the spin-orbit interactions does not affect the degree of demagnetization at long times. Moreover, 
no significant interference effects between the 
laser field and the spin-orbit interactions are seen in $S_z (t)$ and $L_z (t)$.

The inset of Fig.~\ref{fig:figure7} also shows the corresponding demagnetization time scales, which increase from 
$\tau_\text{dm} \simeq 20$~fs for very short pulses ($\tau_p \leq 5$~fs) to $\tau_\text{dm} \simeq 38$~fs for $\tau_p = 
50$~fs. This trend can be qualitatively understood by recalling that longer pulses imply a narrower 
spectral distribution $D_\Psi (\epsilon)$ and thus a slower time evolution of the many-body excited 
state $\ket\Psi$. Moreover, 
in the limit of very short pulses, $\tau_\text{dm} \simeq 20$~fs remains approximately constant 
since after a sudden excitation the demagnetization rate is controlled by the SOC ($\tau_\text{dm} 
\propto \hbar 
/ \xi$, see also Ref.~\onlinecite{Toe15}). 
The results confirm that the laser-induced UFD of TMs reflects primarily the 
intrinsic 
dynamical behavior of the itinerant-electron many-body system, even in cases where the pulse 
duration is longer than 
$\hbar / \xi$ and $\tau_\text{dm}$.

\section{Conclusion}
\label{sec:conclusion}

The laser-triggered dynamics of itinerant-electron magnetism has been investigated in the framework of a many-body 
$pd$ Hamiltonian which describes electron delocalization, Coulomb interactions, spin-orbit interactions and the 
coupling to the laser field on the same footing. The time-dependent many-body state of the system $\ket{\Psi(t)}$ has 
been exactly calculated by applying a numerical short-time Lanczos 
propagation 
method on 
small cluster models with parameters appropriate for Ni. Starting from the ground state $\ket{\Psi_0}$, the time 
evolution of $\ket{\Psi (t)}$ has been followed during and after the laser pulse. The relevant observables, in 
particular the average spin moment $S_z (t)$ and orbital moment $L_z (t)$, have been obtained for a wide range of 
representative excitation parameters: fluence $F$, wave length $\lambda$, linear and circular polarizations 
$\hat \varepsilon$, and pulse duration $\tau_p$. For all considered excitations, cluster models and 
band fillings, 
one observes 
that $S_z(t)$ decreases rapidly after the pulse passage 
reaching values close to its long-time limit $S_z^\infty$ in a very short
characteristic demagnetization time 
$\tau_\text{dm}$ of the order of $20$--$100$~fs.
The actual value of $\tau_\text{dm}$
is found to scale with $\hbar / \xi$, where $\xi$ is the spin-orbit coupling strength, which 
controls the slowest electronic spin-to-orbital angular-momentum transfer. Furthermore, the 
observed general trends show that whenever the main 
ingredients of itinerant-electron magnetism are present, namely, band formation, strong intra-atomic $3d$ Coulomb 
interactions and spin-orbit coupling, the ultrafast demagnetization effect should take place. One 
concludes that the ultrafast demagnetization of ferromagnetic TMs reflects the intrinsic 
many-body dynamical behavior of itinerant magnetism. The universality of 
the effect has been theoretically demonstrated.

The present investigations indicate that ultrafast demagnetization can be regarded as an essentially 
local process which involves 
mainly the atomic spin and orbital $d$-electron degrees of freedom and their immediate local environment. While this 
justifies 
small-cluster modelizations, it is also clear that one would like to improve on this limitation by considering larger 
clusters and extended systems, not least in order to quantify the importance of 
intermediate- 
and long-range dynamical effects. Besides the possible consequences on the electronic correlations, improving on the 
cluster model would allow us to obtain a more quantitative account of the laser absorption efficiency, which has 
been shown to be 
crucial for predicting $S_z(t)$. Such improvements will most certainly involve mean-field or 
functional-integral static approximations of the Coulomb interactions, whose validity could be 
checked by comparison with the exact results 
reported in this work. Moreover, our study suggests that the laser-induced
ultrafast demagnetization effect, being 
an essentially local 
phenomenon, should also take place in ferromagnetic
small clusters, nanoparticles and granular systems. It 
would be therefore most interesting to perform 
cluster-specific studies of ultrafast demagnetization 
in order to reveal its size and structural dependence.

Finally, from the fundamental perspective of understanding the underlying physical mechanisms of 
UFD, it is important to recall that there are other 
forms of spin-lattice relaxations (e.g., electron-phonon coupling) which have been ignored in the present 
electronic model and which are expected to contribute to the magnetization 
dynamics.\cite{Koo10,Koo05,Koo05-PRL,Ste09,Ste10,Fae11,Car11,Mue13}
It would be therefore very interesting to incorporate these contributions into the present 
many-body model, in order to quantify their role at the 
same level as the spin-orbit, Coulomb and hopping electronic effects. 
In addition, other 
excitation methods, for instance, involving hot electron injection, mixed thermalized states and indirect optical 
excitation, 
should also be investigated in order to 
challenge the reliability of the present model and the universal character of the ultrafast demagnetization 
effect.

%%
%\begin{acknowledgments}
%%
%One of the authors (W. T.) acknowledges with thanks the financial support of the Otto-Braun 
%Foundation. Computer 
%resources were provided by the Deutsche Forschungsgemeinschaft and the University of Kassel.
%%
%\end{acknowledgments}
%%

%appendix*
\appendix

%\newpage

%
\section{Interatomic hopping integrals}
\label{app:hopping}

The present Appendix describes the interatomic hopping integrals $t_{jk}^{\alpha \beta}$, which define
the single-particle operator $\hat H_0$ and its band structure as given by 
Eq.~\eqref{eq:tightbinding}. 
The matrix elements $t_{jk}^{\alpha\beta}$ are determined by applying the 
two-center approximation,\cite{Sla54} which has found countless successful applications in the description of the 
electronic structure of solids.\cite{Har89} Since only the $3d$ and $4p$ valence bands are taken into account in the 
model, all hopping elements $t_{jk}^{\alpha \beta}$ in $\hat H_0$ are obtained in terms of the $7$ independent 
Slater-Koster parameters $(dd\sigma)$, $(dd\pi)$, $(dd\delta)$, $(pp\sigma)$, $(pp\pi)$, $(pd\sigma)$ and $(pd\pi)$. 
The corresponding expressions for $t_{jk}^{\alpha \beta}$ are\cite{Sla54}
\begin{equation}%\label{eq:hoppings_all}
  \small
  \begin{split}
    t_{jk}^{4p0,4p0}       ~ &= ~ \lambda_z^2 \, (pp\sigma) \, + \, (1 - \lambda_z^2) \, (pp\pi) \, ,\\	
    t_{jk}^{4p0,4p\pm1}    ~ &= ~ \mp \lambda_z (\lambda_x \pm i\lambda_y) \, [ (pp\sigma) - (pp\pi)] / \sqrt{2} \, ,\\
    t_{jk}^{4p\pm1,4p\pm1} ~ &= ~ [ (1 -\lambda_z^2) \, (pp\sigma) \, + \, (1 + \lambda_z^2) \, (pp\pi)] / 2 \, ,\\
    t_{jk}^{4p+1,4p-1}     ~ &= ~ -(\lambda_x - i\lambda_y)^2 \, [ (pp\sigma) \, - \, (pp\pi) ] / 2 \, ,\\
    t_{jk}^{4p0,3d0}       ~ &= ~ \lambda_z \, [ (3 \lambda_z^2 - 1) \, (pd\sigma) \, + \, 2 \sqrt{3} \, 
(1 - \lambda_z^2) \, (pd\pi) ] / 2 \, ,\\
    t_{jk}^{4p0,3d\pm1}    ~ &= ~ \mp (\lambda_x \pm i \lambda_y) \, [ \sqrt{3} \, \lambda_z^2 \, 
(pd\sigma) \, + \, (1 - 2 \lambda_z^2) \, (pd\pi) ] / \sqrt{2} \, ,\\
    t_{jk}^{4p0,3d\pm2}    ~ &= ~ \lambda_z (\lambda_x \pm i \lambda_y)^2 \, [ \sqrt{3} \, (pd\sigma) 
\, - \, 2 \, (pd\pi) ] / \sqrt{8} \, ,\\
    t_{jk}^{4p\pm1,3d0}    ~ &= ~ \mp (\lambda_x \mp i \lambda_y) \, [ (3 \lambda_z^2 - 1) \, 
(pd\sigma) \, - \, 2 \sqrt{3} \, \lambda_z^2 \, (pd\pi) ] / \sqrt{8} \, ,\\
    t_{jk}^{4p\pm1,3d\pm1} ~ &= ~ \lambda_z \, [ \sqrt{3} \, (1 - \lambda_z^2) \, (pd\sigma) \, + \, 2 
\lambda_z^2 \, (pd\pi) ] / 2 \, ,\\
    t_{jk}^{4p\pm1,3d\pm2} ~ &= ~ \mp (\lambda_x \pm i\lambda_y) \, [ \sqrt{3} \, (1 - \lambda_z^2) \, 
(pd\sigma) \, + \, 2 (1 + \lambda_z^2) \, (pd\pi) ] / 4 \, ,\\
    t_{jk}^{4p\pm1,3d\mp1} ~ &= ~ -\lambda_z (\lambda_x \mp i \lambda_y)^2 \, [ \sqrt{3} \, (pd\sigma) 
\, - \, 2 \, (pd\pi) ] / 2 \, ,\\
    t_{jk}^{4p\pm1,3d\mp2} ~ &= ~ \mp (\lambda_x \mp i \lambda_y)^3 \, [ \sqrt{3} \, (pd\sigma) \, - 
\, 2 \, (pd\pi) ] / 4 \, ,\\
    t_{jk}^{3d0,3d0}       ~ &= ~ (3\lambda_z^2 - 1)^2 \, (dd\sigma) / 4 \, + \, 3 \lambda_z^2 (1 - 
\lambda_z^2) \, (dd\pi) \, + \, 3 (1 - \lambda_z^2)^2 \, (dd\delta) / 4 \, ,\\
    t_{jk}^{3d0,3d\pm1}    ~ &= ~ \mp \lambda_z \sqrt{3} \, (\lambda_x \pm i\lambda_y) \, [ (3\lambda_z^2 - 1) \, 
(dd\sigma) \, + \, 2 (1 - 2\lambda_z^2) \, (dd\pi) \, - \, (1 - \lambda_z^2) \, (dd\delta) ] / \sqrt{8} \, ,\\
    t_{jk}^{3d0,3d\pm2}    ~ &= ~ \sqrt{3} \, (\lambda_x \pm i\lambda_y)^2 \, [ (3\lambda_z^2 - 
1) \, (dd\sigma) \, - \, 4 \lambda_z^2 \, (dd\pi) \, + \, (1 + \lambda_z^2) \, (dd\delta) ] / \sqrt{32} \, ,\\
    t_{jk}^{3d\pm1,3d\pm1} ~ &= ~ [ 3 \lambda_z^2 (1 - \lambda_z^2) \, (d, d; \sigma) \, + \, (4 \lambda_z^4 - 3 
\lambda_z^2 +1) \, (dd\pi) \, + \, (1 - \lambda_z^4) \, (dd\delta) ] / 2 \, ,\\
    t_{jk}^{3d\pm1,3d\pm2} ~ &= ~ \mp \lambda_z (\lambda_x \pm i\lambda_y) \, [ 3 (1 - \lambda_z^2) \, 
(dd\sigma) \, + \, 4 \lambda_z^2 \, (dd\pi) \, - \, (3 + \lambda_z^2) \, (dd\delta) ] / 4 \, ,\\
    t_{jk}^{3d+1,3d-1}     ~ &= ~ -(\lambda_x - i\lambda_y)^2 \, [ 3 \lambda_z^2 \, (dd\sigma) \, + \, 
(1 - 4 \lambda_z^2) \, (dd\pi) \, + \, (\lambda_z^2 - 1) \, (dd\delta) ] / 2 \, ,\\
    t_{jk}^{3d\pm1,3d\mp2} ~ &= ~ \mp \lambda_z (\lambda_x \mp i\lambda_y)^3 \, [ 3 \, (dd\sigma) \, - 
\, 4 \, (dd\pi) \, + \, (dd\delta) ] / 4 \, ,\\
    t_{jk}^{3d\pm2,3d\pm2} ~ &= ~ [ 3 (1 - \lambda_z^2)^2 \, (dd\sigma) \, + \, 4 (1 - \lambda_z^4) \, (dd\pi) \, + \, 
(\lambda_z^4 + 6\lambda_z^2 + 1) \, (dd\delta) ] / 8 \, , ~\text{and}\\
    t_{jk}^{3d+2,3d-2}     ~ &= ~ (\lambda_x - i\lambda_y)^4 \, [ 3 \, (dd\sigma) \, - \, 4 \, (dd\pi) \, + \, 
(dd\delta) ] / 8  \, ,  
  \end{split}
\end{equation}
where $\hat z$ has been chosen as the $m$-quantization axis, and $\lambda_\mu = {\vec R_{jk} \cdot \hat \mu}/{R_{jk}}$ 
denotes the direction cosine of the interatomic vector $\vec R_{jk} = \vec R_j - \vec R_k$ ($\hat \mu = \hat x, \hat y$ 
or $\hat z$). Notice that the 
hopping elements which are not explicitly given can be obtained by applying the relation 
$t_{jk}^{\alpha \beta} = t^{\alpha \beta} (\vec R_{jk}) = [ t^{\beta \alpha} (-\vec R_{jk}) ]^*$. Further details may 
be found in Ref.~\onlinecite{Sla54}.

As described in Sec.~\ref{sec:application}, the $pd$ model has been simplified in order to keep
the dimension of the many-body Hilbert space and the 
numerical effort 
involved in the exact time evolution tractable. Thus, the $3d$ orbitals are approximated by three 
degenerate levels having $|m| \leq 1$ and the $4p$ orbitals by a single level having $m=0$. This reduction of the number 
of bands implies 
that only the four Slater-Koster parameters $(dd\sigma)$, $(dd\pi)$, $(pp\sigma)$ and 
$(pd\sigma)$ are necessary in 
order to determine all hopping integrals $t_{jk}^{\alpha \beta}$.
The corresponding expressions for $t_{jk}^{\alpha \beta}$ are
\begin{equation}\label{eq:hoppings_simplified}
  \small
  \begin{split}
    t_{jk}^{4p0,4p0}       ~ &= ~ (pp\sigma)  \, ,\\
    t_{jk}^{4p0,3d0}       ~ &= ~ \lambda_z \, (pd\sigma)  \, ,\\
    t_{jk}^{4p0,3d\pm1}    ~ &= ~ \mp (\lambda_x \pm i\lambda_y) \, (pd\sigma) / \sqrt{2}  \, ,\\
    t_{jk}^{3d0,3d0}       ~ &= ~ \lambda_z^2 \, (dd\sigma) \, + \, (1 - \lambda_z^2) \, (dd\pi)  \, ,\\
    t_{jk}^{3d0,3d\pm1}    ~ &= ~ \mp \lambda_z (\lambda_x \pm i\lambda_y) \, [ (dd\sigma) - (dd\pi) ] / \sqrt{2}  \, 
,\\
    t_{jk}^{3d\pm1,3d\pm1} ~ &= ~ [ (1 -\lambda_z^2) \, (dd\sigma) \, + \, (1 + \lambda_z^2) \, (dd\pi) ] / 2  \, , ~ 
\text{and}\\
    t_{jk}^{3d+1,3d-1}     ~ &= ~ -(\lambda_x - i\lambda_y)^2 \, [ (dd\sigma) \, - \, (dd\pi) ] / 2 ~.
  \end{split}
\end{equation}
The values of the Slater-Koster parameters are given in the main text.

\section{Electric dipole matrix elements}
\label{app:dipole}

The dominant intra-atomic dipole matrix elements $\bra{\alpha} \vec r \ket{\beta}$ characterizing the interaction $\hat 
H_E$ with the laser field [see Eqs.~\eqref{eq:dipole-approx} and~\eqref{eq:el-field-ham-circular}] can be expressed in 
terms of the irreducible spherical tensor operator 
$\hat T_q^{(k)}$ of rank $k=1$ and components $q$ given by
\begin{equation}\label{eq:tensor}
  \begin{split}
    \hat T_{+1}^{(1)} &= - \dfrac{1}{\sqrt{2}} (\hat x + i\hat y) ~, \\
    \hat T_{-1}^{(1)} &= \dfrac{1}{\sqrt{2}} (\hat x - i\hat y) ~, \\
    \hat T_{0}^{(1)} &= \hat z ~.
  \end{split}
\end{equation}
The elements of $\hat T_q^{(1)}$ between the atomic orbitals $\ket{nlm}$ having principal quantum number $n$, orbital 
angular momentum $l$ and $z$-axis projection $m$ are given by the Wigner-Eckart 
relation\cite{Messiah}
\begin{equation}\label{eq:wet}
  \bra{nlm} \hat T_q^{(k)} \ket{n'l'm'} ~=~ \skp{l'k;m'q}{lm} \, \dfrac{\bra{nl} |\hat T^{(k)}|\ket{n'l'}}{\sqrt{2l' 
  + 1}} ~,
\end{equation}
where $\bra{nl} |\hat T^{(k)}|\ket{n'l'}$ is the reduced matrix element and the scalar products $\skp{l'k;m'q}{lm}$ are 
the Clebsch-Gordan coefficients. Thus, the matrix element $\bra{nlm} \hat T_q^{(k)} \ket{n'l'm'}$ in 
Eq.~\eqref{eq:wet} can be interpreted as a projection resulting from the addition of the angular 
momenta $\vec l'$ and $\vec k$ to $\vec l$ ($\vec l' \oplus \vec k = \vec l$). Since $\bra{nl} |\hat 
T^{(1)}|\ket{n'l'}$ is independent of $m'$, $q$ and $m$, all the dipole matrix elements entering $\hat H_E$ are 
characterized by a single parameter $\bra{3d} |\hat T^{(1)}|\ket{4p}$, the dependence on $m$, $m'$ and $q$ being 
given by the known Clebsch-Gordan coefficients. Since the operator $\hat H_E$ is given by a product of $\hat{\vec r}$ 
and $\vec E$, the matrix element $\bra{3d} |\hat T^{(1)}|\ket{4p}$ gives a measure of the strength of the coupling 
between the electronic translational degrees of freedom and the external electric field $\vec E$ [see 
Eqs.~\eqref{eq:dipole-approx} and \eqref{eq:el-field-ham-circular} of the main text].

The non-vanishing dipole matrix elements then read
\begin{equation}\label{eq:dipole_all}
  \small
  \begin{split}
    \bra{3dm} \hat x \ket{4pm'} ~ &= ~ \left( \delta_{m,m'-1} - \delta_{m,m'+1} \right) \, \left( 
\sqrt{|m|/12} + \delta_{m,0}/6 \right) \, \bra{3d} |\hat T^{(1)}|\ket{4p} ~,\\
    \bra{3dm} \hat y \ket{4pm'} ~ &= ~ i \left( \delta_{m,m'-1} + \delta_{m,m'+1} \right) \, \left( 
\sqrt{|m|/12} + \delta_{m,0}/6 \right) \, \bra{3d} |\hat T^{(1)}|\ket{4p} ~,\\
    \bra{3dm} \hat z \ket{4pm'} ~ &= ~ \delta_{m,m'} \left( |m|/\sqrt{6} + \delta_{m,0} \sqrt{2/9} 
\right) \, \bra{3d} |\hat T^{(1)}|\ket{4p} ~.
  \end{split}
\end{equation}
In the case of circular polarization $\hat \varepsilon_\pm$ the relevant matrix elements are
\begin{equation}\label{eq:sigma1}
   \small
   \begin{split}
   \bra{4pm'} \hat \varepsilon_{\pm} \cdot \hat{\vec r} \ket{3dm} \,&=\, \pm \delta_{m',m \pm 1} \, 
\left( \sqrt{|m|/6} + \sqrt{2} \delta_{m,0} / 6 \right) \, \bra{3d} |\hat T^{(1)}|\ket{4p}^* ~.
    \end{split}
\end{equation}

Taking into account the reduction of the local orbital degeneracy introduced in 
Sec.~\ref{sec:application},
the non-vanishing electric-dipole matrix elements are simplified as follows. For linear 
electric-field polarization, they read
\begin{equation}\label{eq:dipole_simplified}
  \begin{split}
    \bra{3dm} \hat x \ket{4p0} ~ &= ~ -m\, \bra{3d} |\hat T^{(1)}|\ket{4p} / \sqrt{2} ~,\\
    \bra{3dm} \hat y \ket{4p0} ~ &= ~ i (1- \delta_{m,0}) \, \bra{3d} |\hat T^{(1)}|\ket{4p} / \sqrt{2} ~,\\
    \bra{3dm} \hat z \ket{4p0} ~ &= ~ \delta_{m,0} \, \bra{3d} |\hat T^{(1)}|\ket{4p} ~,
  \end{split}
\end{equation}
while for circular polarization they are given by
\begin{equation}\label{eq:sigma1_circular}
   \bra{4p0} \hat \varepsilon_{\pm} \cdot \hat{\vec r} \ket{3dm} \,=\, \pm \delta_{m,\mp1} \, 
\bra{3d} |\hat 
T^{(1)}|\ket{4p}^* ~.
\end{equation}
The value of $\bra{3d} |\hat T^{(1)}|\ket{4p}$ is given in the main text.


\begin{thebibliography}{99}
\bibitem{footnote-fit}
In practice, nearly the same fits are obtained by replacing $S_z(0)$ with the ground-state spin 
magnetization $S_z^0$, 
except when $\tau_p$ is comparable or larger than $\tau_\text{dm}$ as discussed in 
Sec.~\ref{subsec:duration}.
\bibitem{footnote-polarizations}
In addition, other different linear electric field polarizations $\hat \varepsilon$ have also been 
considered. For all $\hat \varepsilon$ 
within the triangle plane the optical absorptions and subsequent dynamics have been found to be 
quantitatively very similar. However, for $\hat \varepsilon$ normal to the triangle plane ($\hat 
\varepsilon = \hat z$) we obtain a strong reduction of the optical absorption. Thus, for a given 
$F$, the demagnetization strength is significantly smaller than 
in the case of in-plane polarizations.
\end{thebibliography}
\end{document}